\documentclass[twocolumn,prd,aps,nofootinbib,showpacs,superscriptaddress]{revtex4-1}
\usepackage{longtable}
\usepackage{multirow}
\usepackage{graphicx}
\usepackage[dvips]{color}
\usepackage{amssymb,amsmath}
\usepackage{supertabular}
\allowdisplaybreaks[4]
\begin{document}

\title{$I=0$ $\pi\pi$ $s$-wave scattering length from lattice QCD}

\author{Ziwen Fu}
\email{fuziwen@scu.edu.cn}
\affiliation{
Key Laboratory for Radiation Physics and Technology of  Education Ministry;
Institute of Nuclear Science and Technology, Sichuan University, Chengdu 610064, China
}

\affiliation{
Center for Theoretical Physics, College of Physical Science and Technology, Sichuan University,
Chengdu 610064, China
}

\author{Xu Chen}
\email{chenxu@scu.edu.cn}
\affiliation{
Key Laboratory for Radiation Physics and Technology of  Education Ministry;
Institute of Nuclear Science and Technology, Sichuan University, Chengdu 610064, China
}

\date{\today}

\begin{abstract}
We deliver lattice results for the $I=0$ $\pi\pi$ elastic $s$-wave scattering length
calculated with the MILC $N_f=3$ flavors of the Asqtad-improved staggered fermions.
The scattering phase shifts are determined by L\"uscher's formula from the energy-eigenvalues of
$\pi\pi$ systems at one center of mass frame and four moving frames
using the moving wall source technique.
Our measurements are good enough to resolve
the scattering length $a$ and  effective range $r$,
moreover, it allows us to roughly estimate the shape parameter $P$.
Using our lattice results, the scattering length $a$ and  effective range $r$  at the physical point
are extrapolated by chiral perturbation theory.
Our results are reasonably consistent with the Roy equation determinations and the newer experimental data.
Numerical computations are carried out with two MILC fine
($a\approx0.09$~fm, $L^3 \times T = 40^3\times 96$)
and one MILC superfine ($a\approx0.06$~fm, $L^3 \times T = 48^3\times 144$) lattice ensembles
at three pion masses of $m_\pi\sim247~{\rm MeV}$, $249~{\rm MeV}$,
and $314~{\rm MeV}$, respectively.
\end{abstract}

\maketitle

\section{Introduction}
The pion-pion ($\pi\pi$) scattering amplitudes are solely predicted at leading order (LO)
in chiral perturbation theory ($\chi$PT)~\cite{Weinberg:1966kf}.
The next-to-leading order (NLO) and next-to-next-to-leading order (NNLO) corrections in the chiral expansion
result in the perturbative deviations from the LO prediction for small pion masses,
and include both estimable non-analytical contributions and
analytical terms with low-energy constants (LEC's)~\cite{Gasser:1983yg,Gasser:1987ah,Bijnens:1995yn,Bijnens:1997vq,Colangelo:2001df},
which can be secured from the experimental measurements or lattice calculations.

With stringent $\chi$PT constraints,
the E865 secured $\pi\pi$ scattering lengths from
the semileptonic $K_{e4}$ decay~\cite{Pislak:2003sv}.
Using Roy equations after the correction for the isospin breaking mass effects,
the NA48/2 decisive analyses of the $K_{e4}$ and
$K_{3\pi}$ decays lead to the robust results
on the $s$-wave $\pi\pi$ scattering lengths~\cite{Batley:2010zza}.
All of these values can be used to inversely determine the important values of the LEC's.

With huge progress of numerical algorithms,
assisted by giant upgrade of computer power,
lattice simulations are able to determine the LEC's values in
the isospin-$2$ $\pi\pi$ scattering with robust statistics~\cite{Beane:2011sc,Dudek:2012gj,Sasaki:2013vxa,Helmes:2015gla}.
Using the  LEC's obtained at the nonphysical pion masses,
the scattering parameters at the physical point can be predicted under the guidance of $\chi$PT.
Moreover, Roy-equation~\cite{Roy:1971tc,Basdevant:1973ru,Ananthanarayan:2000ht}
can determine $\pi\pi$ scattering parameters with trustworthy precision~\cite{Bijnens:1997vq,Colangelo:2001df,GarciaMartin:2011cn},
which can be employed to compare with the relevant lattice evaluations.
NPLQCD Collaboration elegantly demonstrated these strategies  in
the isospin-$2$ $\pi\pi$ scattering~\cite{Beane:2011sc}.

It is well-known that the $I=0$ $\pi\pi$ channel harbors
the lowest resonance: the scalar $\sigma$ or $f_0(500)$ meson,
which have been recently reconfirmed with dispersive analyses and new experimental data~\cite{Pelaez:2015qba}.
Hence, it is highly desirable to investigate the isospin-$0$ $\pi\pi$ interaction properties
directly from lattice QCD.

However, only a few of lattice studies in the isospin-$0$ channel are reported so far.
Kuramashi {\it et~al.} carried out a pioneering work for the isospin-$0$ channel with moving wall source technique~\cite{Kuramashi:1993ka,Fukugita:1994ve}.
A first attempt to extract the sigma from lattice was made in Refs.~\cite{Doring:2011nd,Doring:2011ip}.
With vacuum diagram, we attempted to crudely
calculate the isospin-$0$ $\pi\pi$ scattering,
and estimated the value of  ${\ell_{\pi\pi}^{I=0}}$,
which is a LEC appearing in the NLO $\chi$PT expression
of $I=0$ $\pi\pi$ scattering length~\cite{Fu:2011bz,Fu:2012gf,Fu:2013ffa}.
Liu {\it et~al.} study the isospin-$0$ $\pi\pi$ $s$-wave scattering length
from twisted mass lattice QCD, and evaluate ${\ell_{\pi\pi}^{I=0}}$~\cite{Liu:2016cba}.
A chiral extrapolation of the Hadron Spectrum results is performed
in the isoscalar sector~\cite{Doring:2016bdr}.
Recently, R.~A.~Briceno investigated the isoscalar $\pi\pi$ scattering
and $\sigma$ meson resonance~\cite{Briceno:2016mjc}.
Guo {\it et~al.} studied the elastic phase-shifts for $\pi\pi$
scattering in the scalar and isoscalar channel~\cite{Guo:2018zss}.

It is very impressive that, as it is demonstrated
for the isospin-$2$ $\pi\pi$ scattering~\cite{Beane:2011sc},
the near threshold behavior of the inverse partial wave amplitude
can be exploited to determine the scattering length, effective range and shape parameters,
which can be written in terms of a series of threshold parameters
which satisfy low-energy theorems imposed by chiral symmetry£¬
and encrypt the chief momentum dependence of partial wave amplitude~\cite{Bijnens:1997vq,Beane:2011sc}.
Thus, one can predict the scattering parameters at the physical point
by relevant NLO $\chi$PT expressions~\cite{Beane:2011sc}.
Lattice predictions of scattering parameters are observed
to be  consistent with Roy equation determinations~\cite{Beane:2011sc}.
In this work, we will extend this technique
to the isospin-$0$ $\pi\pi$ scattering.
As it is shown later, this is not a trivial development
since it exhibits some novel features.

Additionally, this approach is only valid in the elastic region,
so the pion mass values should be small enough to be below threshold
where $\sigma$ meson approaches stable~\cite{Liu:2016cba}.
Although the strict value of this threshold is not clear,
the one-loop inverse amplitude method~\cite{Hanhart:2008mx}
indicates that the threshold $m_\pi< 400\, \rm MeV$ should be secure
(see more details in Refs.\cite{Hanhart:2014ssa,Albaladejo:2012te,Pelaez:2010fj}).
In this work, our lattice studies are calculated
at pion masses:  $247\, \rm MeV$, $249\, \rm MeV$ and $314\, \rm MeV$, respectively.
Since pion mass values are all below this threshold,
the influence of the $\sigma$ meson can be reasonably ignored~\cite{Liu:2016cba}.
Therefore, we will only use  $\pi\pi$ correlator
to calculate the $I=0$ $\pi\pi$ elastic $s$-wave scattering lengths,
as it is done in Ref.~\cite{Liu:2016cba}.

In the present study, we will exploit two MILC fine
($a\approx0.09$~fm, $L^3 \times T = 40^3\times 96$),
and one MILC superfine ($a\approx0.06$~fm, $L^3 \times T = 48^3\times 144$)
lattice ensembles with $N_f = 3$ flavors of Asqtad-improved staggered
dynamical quarks~\cite{Aubin:2004wf,Bernard:2001av,stag_fermion}
to compute the isospin-$0$ $s$-wave $\pi\pi$ scattering length,
where L\"uscher's technique~\cite{Luscher:1986pf,Luscher:1990ux,Luscher:1990ck} and its extension\cite{Beane:2003da,Rummukainen:1995vs,Kim:2005gf,Christ:2005gi,Doring:2012eu,
Fu:2011xz,Leskovec:2012gb,Doring:2012eu} are employed to extract
the scattering phases with lattice-calculated energy eigenstates.
The moving wall source technique~\cite{Kuramashi:1993ka,Fukugita:1994ve,Fu:2011wc}
is utilized to calculate the four diagrams classified
for the $I=0$ $\pi\pi$ scattering~\cite{Sharpe:1992pp,Kuramashi:1993ka,Fukugita:1994ve}.

According to the rule of thumb discussions~\cite{Lepage:1989hd,Fu:2016itp},
if we use the fine gauge configurations,
employ lattice ensembles with relatively large spatial dimensions $L$,
and sum the correlators over all time slices ($96$ or $144$),
the signals are anticipated to be significantly improved~\cite{Fu:2016itp}.
Consequently, the signals of vacuum diagram are found to be remarkably improved
as compared with our previous works in Refs.~\cite{Fu:2011bz,Fu:2013ffa}.
It allows us to not only measure the scattering length,
but also explore the effective range.
The chiral extrapolations of the scattering length $m_\pi a_{0}^{I=0}$ is
performed using NLO $\chi$PT. Extrapolated to the physical value of $m_\pi/f_\pi$,
our final results give rise to
$$
m_\pi a_{0}^{I=0} = 0.217(9)(5), \qquad
{\ell_{\pi\pi}^{I=0}}  = 45.6(7.6)(3.8),
$$
which are in reasonable agreement with the recent experimental and theoretical
determinations as well as the lattice calculations.

Most of all, after the chiral extrapolations of the effective range $m_\pi r$
to the physical point, we can probe its lattice result as
$$
m_\pi r =  -6.07(44)(36),
$$
which is also in fair accordance with the Roy-equation determination~\cite{Bijnens:1997vq,Colangelo:2001df}.

This paper is organized as follows.
The L\"uscher's finite volume method,
lattice setup and the computation of the finite volume spectrum of $\pi\pi$ system
are discussed in Sec.~\ref{Sec:Methods}.
The lattice results are given in Sec.~\ref{sec:pipiscattering},
together with relevant fits, which are employed to determine the threshold parameters and the effective range parameters.
In Sec.~\ref{sec:chiextrap},  a summary of the relevant $\chi$PT formulas at NLO
and the chiral extrapolation of lattice-measured data are provided.
A brief summary and some discussions are shown in Sec.~\ref{sec:discussion}.

\newpage
\section{Finite-volume methods}
\label{Sec:Methods}
In the present study, we will examine the $s$-wave $\pi\pi$ system with
the isospin representation of $(I,I_z)=(0,0)$.
We restrict ourselves to the total
momenta $\mathbf{P}=[0,0,0]$, $[0,0,1]$, $[0,1,1]$, $[1,1,1]$ and $[0,0,2]$,
where the momenta are written in units of $\tfrac{2\pi}{L}$.

\subsection{Center of mass frame}
In the center-of-mass frame, the energy levels of two free pions are provided by
$$
E = 2\sqrt{m_\pi^2+ |{\mathbf p}|^2} ,
$$
where ${\mathbf p}=\tfrac{2\pi}{L}{\mathbf n}$, and ${\mathbf n}\in \mathbb{Z}^3$.
The lowest energy $E$ for ${\mathbf n} \ne 0$ (e.g., $\mathbf{n}=(0,0,1)$)
is not within the elastic region $2m_\pi < E < 4m_\pi$,
or beyond the $t$-channel cut, which starts at ${k^2}={m_\pi^2}$~\cite{Beane:2011sc}.
We should remark at this point that the finite-volume methods are only valid for elastic scattering,
consequently, we are only interested in ${\mathbf n} = 0$ for the current study.

Due to the interaction between two pions, the energy levels of $\pi\pi$ system are shifted
by the hadronic interaction from $E$ to $\overline{E}$,
$$
\overline{E} = 2\sqrt{m_\pi^2 + k^2} , \quad k=\frac{2\pi}{L}q ,
$$
where the dimensionless scattering momentum $q \in \mathbb{R}$.
The most important irreducible representation is $A_1^+$.
It is the L\"uscher formula that relates the energy $\overline{E}$
to the $s$-wave $\pi\pi$ scattering phase
$\delta$~\cite{Luscher:1990ux,Luscher:1990ck},
\begin{equation}
\label{eq:CMF}
k \cot\delta(k)=\frac{2}{L\sqrt{\pi}} {\mathcal{Z}_{00}(1;q^2)} ,
\end{equation}
where the zeta function is formally defined by
\begin{equation}
\label{eq:Zeta00_CM}
\mathcal{Z}_{00}(s;q^2)=\frac{1}{\sqrt{4\pi}}
\sum_{{\mathbf n}\in\mathbb{Z}^3} \frac{1}{\left(|{\mathbf n}|^2-q^2\right)^s}  .
\end{equation}
The zeta function $\mathcal{Z}_{00}(s;q^2)$
can be efficiently evaluated by the method described in Ref.~\cite{Yamazaki:2004qb}.
We notice an equivalent L\"uscher formula
has been recently developed in Ref.~\cite{Doring:2011vk},
where the influence of the $d$-wave mixing with $s$-wave for the boosted case
is found to be small.

The lowest energy levels in the center-of-mass frame are below the threshold (namely, $k^2 < 0$)
due to the attractive interaction in the isospin-$0$ $\pi\pi$ scattering,
as it is already noticed in Refs.~\cite{Kuramashi:1993ka,Fukugita:1994ve,Fu:2011bz,Fu:2013ffa,Liu:2016cba}.
It should be worthwhile to stress that the scattering phase shift $\delta(k)$ in
the continuum is solely denoted for $k^2>0$~\cite{Luscher:1990ux}.
As for the case of $k^2 < 0$, it is usual to usher in a phase $\sigma(k)$,
which is associated with $\delta(k)$ via the analytic continuation of $\tan \sigma(k) =
-i \tan \delta(k)$~\cite{Luscher:1990ux}.
Consequently, in the rest of the analysis, it is convenient to always adopt the notation $k\cot \delta(k)$,
as it is already done in Ref.~\cite{Kuramashi:1993ka,Fukugita:1994ve,Fu:2011bz,Fu:2013ffa,Liu:2016cba}.

\subsection{Moving frame}
Using a moving frame with non-zero total momentum
${\mathbf P}=(2\pi/L){\mathbf d}$, ${\mathbf d}\in\mathbb{Z}^3$,
the energy levels of two free pions 
$$
E_{MF} = \sqrt{m_\pi^2 + |{\mathbf p}_1|^2} +
         \sqrt{m_\pi^2 + |{\mathbf p}_2|^2} ,
$$
where ${\mathbf p}_1$, ${\mathbf p}_2$ denote the three-momenta of the pions,
which obey the periodic boundary condition,
${\mathbf p}_1=\frac{2\pi}{L}{\mathbf n}_1$,
${\mathbf p}_2=\frac{2\pi}{L}{\mathbf n}_2$,
${\mathbf n}_1,{\mathbf n}_2\in \mathbb{Z}^3$,
and total momentum ${\mathbf P}$ meets
$
{\mathbf P} = {\mathbf p}_1 + {\mathbf p}_2$~\cite{Rummukainen:1995vs}.

In the presence of the interaction between two pions, the energy $E_{CM}$  is
\begin{equation}
\label{eq:ECM_q}
E_{CM} = 2\sqrt{m_\pi^2 + k^{2}} , \quad k = \frac{2\pi}{L} q ,
\end{equation}
where the dimensionless momentum $q \in \mathbb{R}$, $k =| {\mathbf p}|$,
and ${\mathbf p}$ are quantized to the values
$
{\mathbf p} =\frac{2\pi}{L}{\mathbf r}, {\mathbf r} \in P_{\mathbf d} ,
$
and the set $P_{\mathbf d}$  is
\begin{equation}
\label{eq:set_Pd_MF}
P_{\mathbf d} = \left\{ {\mathbf r} \left|  {\mathbf r} = \vec{\gamma}^{-1}
\left[{\mathbf n}+ \frac{\mathbf d}{2} \right], \right.  {\mathbf n}\in\mathbb{Z}^3 \right\} ,
\end{equation}
where $\vec{\gamma}^{-1}$ is the inverse Lorentz transformation
operating in the direction of the center-of-mass velocity ${\mathbf v}$,
$
\vec{\gamma}^{-1}{\mathbf p} =
\gamma^{-1}{\mathbf p}_{\parallel}+{\mathbf p}_{\perp} ,
$
where ${\mathbf p}_{\parallel}$ and ${\mathbf p}_{\perp}$ are
the ingredients of ${\mathbf p}$ parallel
and perpendicular to ${\mathbf v}$, respectively.
Using the Lorentz transformation,
the energy $E_{CM}$ is connected to the $E_{MF}$  through
$$E_{CM}^2 = E_{MF}^2-{\mathbf P}^2.$$

The first moving frame (MF1) are taken with ${\mathbf d}={\mathbf e}_3$.
We implemented the second moving frame (MF2) with ${\mathbf d}={\mathbf e}_2+{\mathbf e}_3$.
In order to acquire more eigenenergies, we considered the third moving frame (MF3) with
${\mathbf d}={\mathbf e}_1+{\mathbf e}_2 + {\mathbf e}_3$.
The fourth moving frame (MF4) with ${\mathbf d}={2\mathbf e}_3$
is also taken into account.
For these evaluations,
the scattering phase shifts can be acquired from the total energy of the
two-particle system enclosed in a cubic torus
with total momentum ${\bf P} = {2\pi\over  L}{\bf d}$
by the generalized L\"uscher relation~\cite{Luscher:1986pf,Luscher:1990ux,Luscher:1990ck,Rummukainen:1995vs}
\begin{equation}
k \cot\delta(k)  =  {2\over \gamma L \sqrt{\pi}}\ Z_{00}^{\bf d}(1; q^{2})
\ \ \ ,
\label{eqn:RumGott}
\end{equation}
where the $\gamma$-factor is denoted by $\gamma=E_{MF}/E_{CM}$.
The most important irreducible representation is $A_1^+$, which is relevant for
two-particle $s$-wave scattering states in infinite volume.

The evaluation procedure of zeta functions $\mathcal{Z}_{00}^{{\mathbf d}} (1; q^2)$
is discoursed in Appendix A of Ref.~\cite{Yamazaki:2004qb}.
As a consistency check, we exploit the formula~(5) described in Ref.~\cite{Beane:2011sc} to
calculate the zeta functions $\mathcal{Z}_{00}^{{\mathbf d}} (1; q^2)$ as well,
both methods are found to arrive at the same results.
Moreover, we have verified that two approaches are mathematically equivalent for the choice of $\Lambda=1$ in Ref.~\cite{Beane:2011sc}.

\subsection{$\pi\pi$ correlator function}
In this section, we follow original derivations and notations
in Refs.~\cite{Sharpe:1992pp,Kuramashi:1993ka,Fukugita:1994ve}
to study $\pi\pi$ scattering of two Nambu-Goldstone pions
in Asqtad-improved staggered dynamical fermion formalism.
We build the isospin-$0$ $\pi\pi$ state
with following interpolating operator~\cite{Kuramashi:1993ka,Fukugita:1994ve}~\footnote{
In this operator, the momentum ${\mathbf{ p}}$ is associated
with time $t+1$, and there is another similar operator where
the momentum ${\mathbf{ p}}$ is associated with time $t$.
After averaging over the two operators, we build an operator,
which is symmetric by swapping the momentum ${\mathbf{ 0}}$ and ${\mathbf{ p}}$,
and preserve the parity symmetry.
Please consult the technical issue of constructing a parity conserving operator in Ref.~\cite{Feng:2014gba}.
Note that this parity violation in Eq.~(\ref{EQ:op_pipi}) vanishes
when $t+1$ is replaced by $t$, consequently,
it tends to be a small effect. It should be incorporated into more sophisticated
lattice study in the future.
}
\begin{eqnarray}
\label{EQ:op_pipi}
 {\cal O}_{\pi\pi}^{I=0} (\mathbf{p},t) &=&
 \frac{1}{\sqrt{3}}
 \Bigl\{  \pi^{-}(t) \pi^{+}(\mathbf{p},t+1) +
          \pi^{+}(t) \pi^{-}(\mathbf{p},t+1) \cr
          &&-
          \pi^{0}(t) \pi^{0}(\mathbf{p},t+1) \Bigl\} .
\end{eqnarray}
with the interpolating pion operators denoted by
\begin{eqnarray}
{\pi^+}(t) &=& -\sum_{\bf{x}} \bar{d}({\bf{x}}, t)\gamma_5 u({\bf{x}},t) , \cr
{\pi^-}(t) &=& \sum_{\bf{x}} \bar{u}({\bf{x}}, t)\gamma_5 d({\bf{x}},t), \cr
{\pi^0}(t) &=&
\frac{1}{\sqrt{2}}\sum_{\bf{x}} [
\bar{u}({\bf{x}},t)\gamma_5 u({\bf{x}},t) -
\bar{d}({\bf{x}},t)\gamma_5 d({\bf{x}},t) ] . \nonumber
\end{eqnarray}
Note that the operator ${\cal O}_{\pi\pi}^{I=0} (\mathbf{p},t)$ belongs to $A_1^+$.

The $\pi\pi$ four-point correlation function
with the momentum ${\mathbf p}$ can be expressed as,
\begin{eqnarray}
\label{EQ:4point_pK_mom}
\hspace{-0.5cm}C_{\pi\pi}({\mathbf p}, t_4,t_3,t_2,t_1) &=&
\sum_{\mathbf{x}_1}\sum_{\mathbf{x}_2}\sum_{\mathbf{x}_3}\sum_{\mathbf{x}_4}
e^{ i{\mathbf p} \cdot ({\mathbf{x}}_4 -{\mathbf{x}}_2) } \times\cr
&&\hspace{-1.6cm}\langle
{\cal O}_{\pi}({\bf{x}}_4, t_4)
{\cal O}_{\pi}({\bf{x}}_3, t_3)
{\cal O}_{\pi}^{\dag}({\bf{x}}_2, t_2)
{\cal O}_{\pi}^{\dag}({\bf{x}}_1, t_1) \rangle ,
\end{eqnarray}
where we usually choose $t_1 =0$, $t_2=1$, $t_3=t$,
and $t_4 = t+1$ to prevent the intricate color Fierz rearrangement of the
quark lines~\cite{Kuramashi:1993ka,Fukugita:1994ve} .

\begin{figure}[h!]
\includegraphics[width=8.0cm,clip]{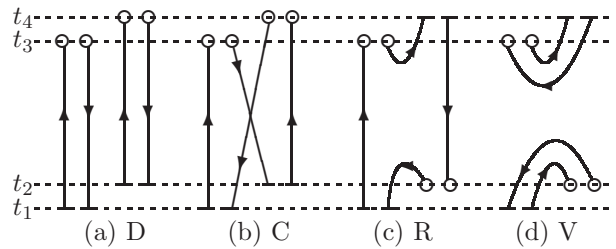}
\caption{ \label{fig:diagram}
Quark-line diagrams for $\pi\pi$ correlation functions.
The wall sources are represented by short bars,
and open circles indicate the wall sinks for local pion operators.
}
\end{figure}

In the isospin limit, $\pi\pi$ scattering amplitude receives only four contributions,
we exhibit these quark-line diagrams in Fig.~\ref{fig:diagram},
which are normally classified as direct ($D$), crossed ($C$), rectangular ($R$),
and vacuum ($V$) diagrams, respectively~\cite{Kuramashi:1993ka,Fukugita:1994ve,Sharpe:1992pp}.

In the present study, we  employ  moving wall source technique
to evaluate four quark-line diagrams~\cite{Kuramashi:1993ka,Fukugita:1994ve}.
In our previous studies~\cite{Fu:2012gf,Fu:2013ffa},
we present a detailed procedure to express
these diagrams in the center-of-mass frame~\cite{Fu:2013ffa}
with the light quark propagator $G$~\cite{Kuramashi:1993ka,Fukugita:1994ve},
and the relevant expressions in the moving frame are also provided in Ref.~\cite{Fu:2012gf}.
To write them in the generic frame ({\it i.e.}, the momenta $\mathbf{p}=[0,0,0]$, $[0,0,1]$, $[0,1,1]$, $[1,1,1]$ and $[0,0,2]$),
we used an up quark source with $1$, and an
anti-up quark source with $e^{i{\mathbf p} \cdot {\mathbf{x}} }$
(except for $V$, where we use $1$)
on each site for two pion creation operator, respectively, then relevant expressions
can be written as
\begin{widetext}
\begin{eqnarray}
\label{eq:dcr}
C^D_{\pi\pi}({\mathbf p},t_4,t_3,t_2,t_1) &=&
\sum_{ \mathbf{x}_3, \mathbf{x}_4} \cos({\mathbf p} \cdot {\mathbf{x}}_4)
\langle \mbox{Tr}
[G_{t_1}^{\dag}({\mathbf{x}}_3,t_3)G_{t_1}({\mathbf{x}}_3,t_3)]
\mbox{Tr}
[G_{t_2}^{\dag}({\mathbf{x}}_4,t_4)G_{t_2}({\mathbf{x}}_4,t_4)] \rangle,\cr
C^C_{\pi\pi}({\mathbf p},t_4,t_3,t_2,t_1) &=&
\sum_{\mathbf{x}_3, \mathbf{x}_4} \cos({\mathbf p} \cdot {\mathbf{x}}_4)
\langle \mbox{Tr}
[G_{t_1}^{\dag}({\mathbf{x}}_3,t_3)G_{t_2}({\mathbf{x}}_3,t_3)
 G_{t_2}^{\dag}({\mathbf{x}}_4,t_4)G_{t_1}({\mathbf{x}}_4,t_4)] \rangle,\cr
C^R_{\pi\pi}({\mathbf p},t_4,t_3,t_2,t_1) &=&
\sum_{\mathbf{x}_2,\mathbf{x}_3}
\cos({\mathbf p} \cdot {\mathbf{x}}_2)
\langle \mbox{Tr}
[G_{t_1}^{\dag}({\mathbf{x}}_2, t_2) G_{t_4}({\mathbf{x}}_2, t_2)
 G_{t_4}^{\dag}({\mathbf{x}}_3, t_3) G_{t_1}({\mathbf{x}}_3, t_3) ] \rangle, \cr
C_{\pi\pi}^V({\mathbf p},t_4,t_3,t_2,t_1) &=&
\sum_{\mathbf{x}_2,\mathbf{x}_3 }
\cos({\mathbf p} \cdot ({\mathbf{x}}_2  - {\mathbf{x}}_3))
\langle
\mbox{Tr} [G_{t_1}^{\dag}(\mathbf{x}_2,t_2)G_{t_1}(\mathbf{x}_2,t_2)]
\mbox{Tr} [G_{t_4}^{\dag}(\mathbf{x}_3,t_3)G_{t_4}(\mathbf{x}_3,t_3)]
\rangle ,  \cr
&& - \delta_{{\mathbf p},{\mathbf 0} }
\sum_{\mathbf{x}_2,\mathbf{x}_3 } \langle \mbox{Tr}
[G_{t_1}^{\dag}({\bf{x}}_2, t_2) G_{t_1}({\bf{x}}_2, t_2) \rangle
\langle \mbox{Tr}[G_{t_4}^{\dag}({\bf{x}}_3, t_3) G_{t_4}({\bf{x}}_3, t_3) ]\rangle ,
\end{eqnarray}
where we note the fact that for the momenta ${\mathbf p} \ne {\mathbf 0}\equiv[0,0,0]$,
\begin{eqnarray}
\sum_{\mathbf{x}_2,\mathbf{x}_3 }
\cos({\mathbf p} \cdot ({\mathbf{x}}_2  - {\mathbf{x}}_3)) \langle \mbox{Tr}
[G_{t_1}^{\dag}({\bf{x}}_2, t_2) G_{t_1}({\bf{x}}_2, t_2) \rangle
\langle \mbox{Tr}[G_{t_4}^{\dag}({\bf{x}}_3, t_3) G_{t_4}({\bf{x}}_3, t_3) ]\rangle  =0 . \nonumber
\end{eqnarray}
\end{widetext}
The combinations of light quark propagator $G$ that we apply for $\pi\pi$ four-point
functions are shown in Fig.~\ref{fig:diagram}.
Note that the vacuum diagram is not accompanied
by a vacuum subtraction for non-zero momenta ${\mathbf p}$~\cite{Fu:2012gf}.

It should be worthwhile to mention that the treatment of vacuum  part
is actually similar to that of the disconnected piece for sigma operator
in our previous study~\cite{Bernard:2007qf},
where Professor Carleton DeTar has especially designed a FFT algorithm to
calculate the disconnected part~\cite{Bernard:2007qf}.
In a same manner, the vacuum diagram can be calculated
with a FFT algorithm, which is courteously
dedicated to  Appendix~\ref{app:FFT}.
In practice, we found that it indeed saves computer resource significantly.

The $\pi\pi$ correlation functions can be expressed in terms of
four diagrams~\cite{Sharpe:1992pp,Kuramashi:1993ka,Fukugita:1994ve},
\begin{eqnarray}
\label{EQ:phy_I0_2}
C_{\pi\pi}^{I=0}(t) \equiv D + \frac{N_f}{2} C - 3N_f R + \frac{3}{2}V ,
\end{eqnarray}
where the staggered-flavor factor $N_f=4$ is introduced due to the number of tastes
natural to the Kogut-Susskind formulation~\cite{Sharpe:1992pp}.
Using the quadruple root of the fermion determinant,
the four-fold degeneracy of the staggered sea quarks
is believed to be neatly removed~\cite{Sharpe:1992pp}.

We should remember that,
the contributions of non-Nambu-Goldstone
pions in the intermediate states is exponentially
suppressed for large $t$~\cite{Sharpe:1992pp,Kuramashi:1993ka,Fukugita:1994ve}.
Hence, we think that $\pi\pi$ interpolator
does not greatly couple to other $\pi\pi$  tastes,
and neglect this systematic errors.

\newpage
\subsection{Lattice Calculation}
We employed the MILC gauge configurations with three Asqtad-improved
staggered sea quarks~\cite{Bernard:2010fr,Bazavov:2009bb}.
The simulation parameters are summarized in Table~\ref{tab:MILC_configs}.
By MILC convention, lattice ensembles are
referred to as ``fine'' for $a\approx0.09$~fm,
and ``super-fine'' for $a\approx0.06$~fm.
For easy notation, it is handy to adopt $(am_l, am_s)$ to categorize lattice ensembles.
We should bear in mind that MILC gauge configurations are
generated using the staggered formulation of lattice
fermions~\cite{Kaplan:1992bt} with the fourth root of
fermion determinant~\cite{Bernard:2001av}.
All the gauge configurations were gauge fixed to Coulomb gauge
before calculating light quark propagators.

\begin{table*}[t!]
\caption{ \label{tab:MILC_configs}
Simulation parameters of the MILC lattice ensembles.
Lattice dimensions are described in lattice units with spatial ($L$) and temporal ($T$) size.
The gauge coupling  $\beta$ is shown in Column $3$.
The fourth block give bare masses of the light and
strange quark masses in terms of $am_l$ and $am_s$, respectively.
Column $5$ gives pion masses in MeV.
The lattice spatial dimension ($L$) in {\rm fm} and in units of the finite-volume pion mass
are given in Column $6$ and $7$ respectively.
We also list the mass ratio $m_\pi/f_\pi$.
The  values of the calculated $\pi\pi$ correlators for each of the lattice ensembles are shown in Column $9$,
and the last Column gives the number of gauge configurations used in this work.
}
\begin{ruledtabular}
\begin{tabular}{llllccllll}
Ensemble &$L^3 \times T$ &$\beta$ & $a m_l/a m_s$
& $m_\pi({\rm MeV})$ &$L{\rm(fm)}$  &$m_\pi L$
& $m_\pi/f_\pi$ & $N_{\rm slice}^{\pi\pi}$ &$N_{\rm cfg}$  \\
\hline
\multicolumn {10}{c}{$a \approx 0.06$~fm}        \\
48144f21b747m0036m018 & $48^3\times144$  & $7.47$  & $0.0036/0.018$   & $314$
                      & $2.8$           & $4.49$   & $2.23(2)$       & $144$   & $102$ \\
\multicolumn {10}{c}{$a \approx 0.09$~fm}        \\
4096f21b708m0031m031  & $40^3\times96$  & $7.08$   & $0.0031/0.031$   & $247$
                      & $3.4$           & $4.21$   & $1.70(1)$       & $96$   & $604$ \\
4096f3b7045m0031      & $40^3\times96$  & $7.045$  & $0.0031/0.0031$  & $249$
                      & $3.4$           & $4.20$   & $1.76(1)$       & $96$   & $560$ \\
\end{tabular}
\end{ruledtabular}
\end{table*}

Although it is expensive,
the moving wall source technique~\cite{Kuramashi:1993ka,Fukugita:1994ve}
is believed to be able to calculate the relevant correlators with high quality.
We extend this method to two-particle system with nonzero momenta
to examine the $\kappa$, $\sigma$,  and $K^\star(892)$ meson decays~\cite{Fu:2011xw,Fu:2012gf,Fu:2012tj},
and meson-meson scattering~\cite{Fu:2011bz,Fu:2013ffa,Fu:2011wc}.
From these works, we found that moving-wall source technique can calculate
four-point correlators with desirable quality.

To compute $\pi\pi$ correlators,
the conjugate gradient method is used to get
the matrix element of the inverse Dirac fermion matrix.
We compute the correlators on all the $T$ time slices,
and explicitly combine theses results,
namely, the correlator $C_{\pi\pi}(t)$ is measured by
\begin{eqnarray}
 C_{\pi\pi}(t) &=&
\frac{1}{T}\sum_{t_s=0}^{T} \left\langle
\left(\pi\pi\right)(t+t_s)\left(\pi\pi\right)^\dag(t_s)\right\rangle .
\nonumber
\end{eqnarray}
After averaging the propagators over all the $T$ values,
the statistics are found to be remarkably improved~\footnote{
For each gauge configuration, we compute $3T=3*96=284$ or $3T=3*144=432$  light quark propagators.
In practice, we save all the light $u$ quark propagators into the moving disks,
when needed, they are copied into computer memory.
Using this strategy, then for each time-slice calculation,
it averagely costs only three light quark propagator computations for three colors.
On this point of view, it is actually ``cheap''.
}.

According to the analytical arguments in Refs.~\cite{Lepage:1989hd,Fukugita:1994ve}
and the semi-empirical discussions in Ref.~\cite{Fu:2016itp},
the noise-to-signal ratio of $\pi\pi$ correlator
is improved approximately  $\propto 1/\sqrt{N_{\rm slice} L^3}$,
where $L$ is the lattice spatial dimension, and $N_{\rm slice}$ is the
number of the time slices calculated the light propagators for a given lattice ensemble.
In the present study, we exploit the MILC lattice ensembles
with the relatively large $L$ (40 or 48),
and sum $\pi\pi$ correlators over all the time slices (96 or 144),
consequently, it is natural that the signals of $\pi\pi$ correlators
should be significantly improved.
Admittedly, the most efficient way to improve the relevant noise-to-signal ratio
is to use the finer gauge configurations or equivalent anisotropic gauge configurations~\cite{Briceno:2016mjc}.

We compute two-point pion correlators with the zero and none-zero
momenta ($\mathbf{0}$ and $\mathbf{p}$) as well,
\begin{eqnarray}
\label{eq:pi_cor_PW_k000}
C_\pi({\mathbf 0}, t) &=& \frac{1}{T}\sum_{t_s=0}^{T-1}
\langle 0|\pi^\dag ({\mathbf 0}, t+t_s) W_\pi({\mathbf 0}, t_s) |0\rangle, \\
\label{eq:pi_cor_PW_k100}
C_\pi({\mathbf p}, t) &=& \frac{1}{T}\sum_{t_s=0}^{T-1}
\langle 0|\pi^\dag ({\mathbf p}, t+t_s) W_\pi({\mathbf p}, t_s) |0\rangle,
\end{eqnarray}
where $\pi$ is pion point-source operator
and $W_\pi$ is pion wall-source operator~\cite{Bernard:2001av,Aubin:2004wf}.
To simplify notations, the summation over lattice space point in sink
is not written out. It should be worthwhile to stress that
the summations over all the time slices for $\pi$ propagators
guarantee the extraction of pion mass with high precision
(see the semi-empirical discussions in Ref.~\cite{Fu:2016itp}).

Discarding the contributions from the excited states,
the pion mass $m_\pi$ and energy $E_\pi({\mathbf p})$
can be robustly extracted at large $t$ from the two-point pion correlators~(\ref{eq:pi_cor_PW_k000})
and (\ref{eq:pi_cor_PW_k100}), respectively~\cite{Bazavov:2009bb},
\begin{eqnarray}
\label{eq:pi_fit_PW_k000}
\hspace{-0.6cm} C_\pi({\mathbf 0}, t) &=& A_\pi(\mathbf{0}) \left[e^{-m_\pi t}+e^{-m_\pi(T-t)}\right] +\cdots, \\
\label{eq:pi_fit_PW_k100}
\hspace{-0.6cm} C_\pi({\mathbf p}, t) &=& A_\pi(\mathbf{p})
\left[e^{-E_\pi({\mathbf p}) t}+e^{-E_\pi({\mathbf p})(T-t)}\right] + \cdots,
\end{eqnarray}
where the ellipses indicate the oscillating parity partners,
and $A_\pi(\mathbf{0})$ and $A_\pi(\mathbf{p})$ are two overlapping amplitudes,
which are used to evaluate the wrap-around contributions
for $\pi\pi$ correlators subsequently~\cite{Gupta:1993rn,Umeda:2007hy,Feng:2009ij}.

Using the method described in MILC's work~\cite{Aubin:2004fs,Bernard:2007ps},
we can calculate the pion decay constants $f_\pi$ for three MILC ensembles.
Our fitted values of the pion decay constants
are found to be in well agreement with the same quantities
which are professionally computed on the same lattice ensembles
by the MILC collaboration~\cite{Aubin:2004fs,Bernard:2007ps,Bazavov:2009bb}.
The MILC¡¯s determinations on pion
decay constants are directly quoted in this work, and
are listed in Table~\ref{tab:MILC_configs} with $m_\pi/f_\pi$.

\subsection{ Extraction of energies }
\label{SubSec:Extraction of energies}

The energy $E_{\pi\pi}$ of $\pi\pi$ system can be extracted  from $\pi\pi$ four-point function as~\cite{Golterman:1985dz,DeTar:2014gla}
\begin{eqnarray}
\label{eq:E_pionpion}
\hspace{-1.0cm} C_{\pi\pi}(t)  &=&
Z_{\pi\pi}\cosh\left[E_{\pi\pi}\left(t - \frac{1}{2}T\right)\right] \cr
&& + (-1)^t Z_{\pi\pi}^{\prime}\cosh \left[E_{\pi\pi}^{\prime} \left(t-\frac{1}{2}T\right)\right] + \cdots
\end{eqnarray}
for a large $t$ to suppress the excited states,
the terms alternating in sign is a symbol of a staggered scheme,
and the ellipsis implies the contributions from the excited states
which are reduced exponentially.
In practice, the pollution due to the ``wraparound'' effects~\cite{Gupta:1993rn,Umeda:2007hy,Feng:2009ij}
should be considered.
Actually, we will withdraw the ``wraparound''  pollution before fitting with this formula,
as practiced in our former works~\cite{Fu:2011bz,Fu:2012gf,Fu:2013ffa}.

\section{FITTING ANALYSES}
\label{sec:pipiscattering}
\subsection{Lattice Phase Shift}

In order for a more intuitive presentation of our lattice results,
we evaluate the ratios~\cite{Kuramashi:1993ka,Fukugita:1994ve},
\begin{equation}
\label{EQ:ratio}
R^X(t) = \frac{ C_{\pi\pi}^X(0,1,t,t+1) }{ C_\pi (0,t) C_\pi(1,t+1) }, \quad  X = D, C, R,  {\rm and} \, V,
\end{equation}
where $C_\pi (0,t)$ and $C_\pi (1,t+1)$ are pion correlators  with
a given momentum~\cite{Kuramashi:1993ka,Fukugita:1994ve}.
We should remark at this point that our visualizations of
various diagrams for the $\pi\pi$ correlator is
somewhat analogous to those of the Hadron Spectrum Collaboration in Ref.~\cite{Briceno:2016mjc},
where the time dependence of  $\pi\pi$ correlator is
weighted by $e^{E_0 t}$ with $E_0$ the energy of the lightest state.
As a matter of fact, both visualized methods indicate the corresponding energy shifts of $\pi\pi$ system.

\begin{figure}[t!]
\includegraphics[width=8.5cm,clip]{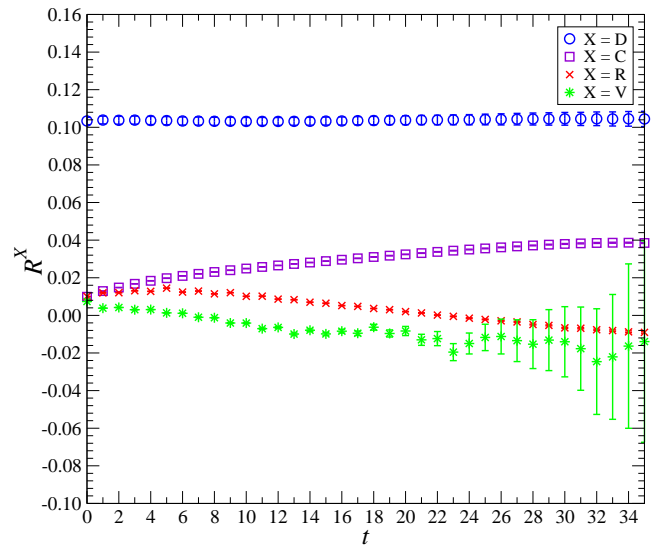}
\caption{\label{fig:rcfv_r_0000_I0} (color online).
Individual amplitude ratios $R^X(t)$ of $\pi\pi$ four-point functions computed
via the moving wall source technique at $\mathbf{P}=[0,0,0]$ for MILC ensemble $(0.0031,0.031)$:
direct diagram (blue circle) displaced by $0.9$, vacuum diagram (green stars),
crossed diagram (magenta squares) and rectangular diagram (red crosses).
}
\end{figure}

In Fig.~\ref{fig:rcfv_r_0000_I0}, we display the various contributions to
an example $\pi\pi$  correlator for a MILC ensemble $(0.0031,0.031)$ at $\mathbf{P}=[0,0,0]$
as the functions of $t$, which are illustrated as the individual ratios $R^X(t), \,\, X = D, C, R,  {\rm and} \, V$.
The vacuum deduction, which is time independent, is carried out with the technique described in Ref.~\cite{Fu:2013ffa}.
It is worth stressing once again that the vacuum diagram is not accompanied by a vacuum subtraction for the cases with the non-zero momenta~\cite{Fu:2012gf}.

The ratio values of the direct amplitude $R^D$ are quite close to
oneness~\cite{Kuramashi:1993ka,Fukugita:1994ve,Fu:2011bz,Fu:2012gf,Fu:2013ffa,Liu:2016cba}.
In fact, the $I=0$ $\pi\pi$  scattering  is perturbative
at low momentum and at small light-quark masses,
as mandated by $\chi$PT.
Consequently, in a finite volume, two-pion energies deviate slightly
from the noninteracting energies; namely, the sum of the pion masses (or the boosted pion masses for moving frames).
Moreover, in lattice practice,  the ratios of these energy shifts to total energies are found normally
to be about $5\%$~\cite{Kuramashi:1993ka,Fukugita:1994ve,Fu:2011bz,Fu:2012gf,Fu:2013ffa,Liu:2016cba}.

The systematically oscillating behavior of rectangular amplitude is clearly observed,
which is a typical feature of the staggered formulation of
lattice fermions~\cite{Golterman:1985dz}.
In this work, we make use of DeTar's strategy~\cite{DeTar:2014gla} (namely, Eq.~(\ref{eq:E_pionpion}))
to remove this oscillating contribution.
For the vacuum amplitude of this MILC lattice ensemble, we nicely obtain a good signal up to $t \sim 26$, beyond which signals are quickly lost (see Fig.~\ref{fig:rcfv_r_0000_I0}).
We notice that all the four diagrams are measured with good signals,
and make a remarkable progress
as compared with our previous works~\cite{Fu:2011bz,Fu:2012gf,Fu:2013ffa}.

\begin{figure}[t!]
\includegraphics[width=8.5cm,clip]{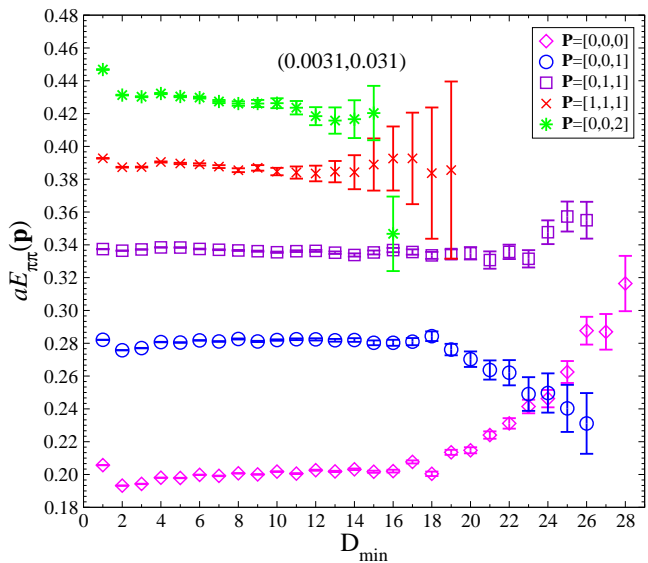}
\caption{\label{fig:eff_eng_I0} (color online).
Effective energy $E_{\pi\pi}$ plots for the MILC ensemble $(0.0031,0.031)$ as the functions of $\rm D_{min}$
for the $\pi\pi$ scattering in the $I=0$ channel in lattice units.
}
\end{figure}

As already explained in previous studies~\cite{Fu:2011bz,Fu:2012gf,Fu:2013ffa},
a persuasive way to process our lattice data is the  resort to the ``effective energy'' plot, which is a variant of the effective mass plot, and  very similar to the
``effective scattering length'' plot~\cite{Beane:2007xs}.
\begin{table*}[th!]	
\caption{\label{tab:pp_Io_kcotk}
Summaries of the lattice results for the fitted energies $E_{\pi\pi}$ of the $I=0$ $\pi\pi$ system.
The third block shows the fitted energies $E_{\pi\pi}$ in lattice units.
Column four gives the fitting range,
and Column five indicates the number of degrees of freedom (dof) for the fit.
The six block is the center-of-mass scattering momentum $k^2$ in terms of $m_\pi^2$,
and Column seven gives the values of $k \cot\delta/m_\pi$, which are calculated
by L\"uscher formula~(\ref{eq:CMF}) or (\ref{eqn:RumGott}).
}
\begin{ruledtabular}
\begin{tabular}{lllllll}
$\rm Ensemble$  & $\mathbf{P}$     & $a E$  &  {\rm Range} & $\chi^2/{\rm dof}$
& $k^2/m_\pi^2$   & $k \cot \delta/m_\pi$   \\
\hline
\multirow{6}*{$(0.0036, 0.018)$} & $[0,0,0]$ & $0.17704(88)$      & $14-28$   & $14.4/11$
                     & $-0.01043(91)$      & $0.638(122)$         \\
          &$[0,0,1]$ & $0.23896(176)$      & $13-24$              & $10.6/8$
                     & $0.1421(241)$       & $0.526(155)$         \\
          &$[0,1,1]$ & $0.28164(230)$      & $13-21$              & $2.4/5$
                     & $0.2874(371)$       & $0.396(146)$         \\
          &$[1,1,1]$ & $0.31862(299)$      & $12-23$              & $16.5/8$
                     & $0.4321(544)$       & $0.449(187)$         \\
          &$[0,0,2]$ & $0.35776(276)$      & $12-25$              & $5.7/10$
                     & $0.6990(564)$       & $0.478(315)$         \\
\hline
\multirow{6}*{$(0.0031, 0.031)$} & $[0,0,0]$ & $0.20207(103)$     & $16-25$     & $12.4/6$
                     & $ -0.0787(77)$      & $1.423(216)$         \\
          &$[0,0,1]$ & $0.28198(136)$      & $14-25$              & $8.9/8$
                     & $0.2373(174)$       & $1.277(240)$         \\
          &$[0,1,1]$ & $0.33510(131)$      & $13-23$              & $9.1/7$
                     & $0.4203(198)$       & $0.919(142)$         \\
          &$[1,1,1]$ & $0.38406(363)$      & $11-19$              & $5.3/5$
                     & $0.6580(629)$       & $1.324(648)$         \\
          &$[0,0,2]$ & $0.42356(403)$      & $11-22$              & $12.1/8$
                     & $0.8211(771)$  & $0.832(723)$              \\
\hline
\multirow{6}*{$(0.031, 0.0031)$} & $[0,0,0]$ & $0.20169(103)$     & $15-27$     & $2.2/9$
                     & $-0.00804(96)$      & $1.382(260)$         \\
          &$[0,0,1]$ & $0.28233(147)$      & $14-24$              & $8.6/7$
                     & $0.2429(188)$       & $1.346(277)$         \\
          &$[0,1,1]$ & $0.33514(199)$      & $12-19$              & $2.2/4$
                     & $0.4236(303)$       & $0.937(224)$         \\
          &$[1,1,1]$ & $0.38130(437)$      & $10-18$              & $10.6/5$
                     & $0.6134(752)$       & $1.038(563)$         \\
          &$[0,0,2]$ & $0.42521(303)$      & $10-24$              & $8.9/11$
                     & $0.8562(583)$       & $1.061(681)$
\end{tabular}
\end{ruledtabular}
\end{table*}
In practice, the isospin-$0$ $\pi\pi$ four-point functions were fit by
altering the minimum fitting distances $\rm D_{min}$,
and putting the maximum distance $\rm D_{max}$ either at $T/2$
or where the fractional statistical errors exceed about $20\%$
for two sequential time slices~\cite{Bernard:2001av}.
The example ``effective energy'' plots for the MILC ensemble $(0.0031,0.031)$
as the functions of $\rm D_{min}$
are illustrated in Fig.~\ref{fig:eff_eng_I0}.
For $\mathbf{P}=[0,0,0]$, the plateau is clearly observed from $\rm D_{min} = 8 \sim 16$.
We also notice that this plateau is distorted starting from $t=17$,
since the relevant vacuum contribution becomes noisy after that,
as compared with the other contributions,
please see Fig.~\ref{fig:rcfv_r_0000_I0} for details.

The energies $a E_{\pi\pi}$ of $\pi\pi$ system
are secured from the ``effective energy'' plots,
and usually chosen by looking for a combination of a ``plateau'' in the
effective energy plots as the function of  $\rm D_{min}$,
a good confidence level, and $\rm D_{min}$ large enough
to suppress the excited states~\cite{Beane:2007xs,Feng:2009ij}.
It should be worthwhile to stress once again that
 the``wraparound''  contamination~\cite{Gupta:1993rn,Umeda:2007hy,Feng:2009ij}
are removed before fitting by subtracting  the lattice-calculated wraparound contribution
from the relevant correlators.
In the present study, our measured quantities from two-point
functions are sufficiently precise to allow us to subtract the
wraparound contributions, and the unwanted finite-T effects
are anticipated to be neatly removed.
The details of how to calculate the ``wraparound''  contamination in
center-of-mass frame and the moving frames are provided in our previous work~\cite{Fu:2016itp}.

The fitted values of the energies $a E_{\pi\pi}$ of $\pi\pi$ system,
fit range and fit quality ($\chi^2/{\rm dof}$)
are given in Table~\ref{tab:pp_Io_kcotk}.
The fit qualities $\chi^2/{\rm dof}$ are turned out to be reasonable.
Now it is straightforward to substitute these fitted energies $a E_{\pi\pi}$
into L\"uscher formula~(\ref{eq:CMF}) or (\ref{eqn:RumGott})
to secure the relevant values of the $k \cot \delta/m_\pi$.
All of these values are also summarized in Table~\ref{tab:pp_Io_kcotk},
where the statistical errors of $k^2$ are calculated from
both the statistical errors of the energies $a E_{\pi\pi}$ and pion masses $am_\pi$.
Note that the relevant scattering momenta $k$ are cited in units of
pion masses $m_\pi$ in order to explicate the remaining analysis
in such a manner which is independent of the scale setting,
as it is suggested by NPLQCD Collaboration in Ref.~\cite{Beane:2011sc}.
All of these five states (namely, $\mathbf{P}=[0,0,0]$, $[0,0,1]$, $[0,1,1]$, $[1,1,1]$ and $[0,0,2]$ ) for each lattice ensemble
will been analyzed to produce the effective range expansion parameters.

\subsection{The effective range approximation parameters}
\label{se:erap}
As it is shown latter, the effective range approximation is an expansion of the real part
of the inverse partial wave scattering amplitude $[\mbox{Re}\, t(k)]^{-1}$ in powers of $k$,
the magnitude of the center-of-mass three-momentum of each pion~\cite{Beane:2011sc},~\footnote{
In Ref.~\cite{Adhikari:1983ii}, the effective range function $k^{-1} \tan\delta$
is used instead of the effective range function $k \cot\delta$,
due to the occurrence of subthreshold poles in $k \cot\delta$.
Moreover, the regime of validity for the effective range expansion is
discussed in Ref.~\cite{Beane:2003da}.
}
namely,
\begin{equation}
\frac{k\,\cot{\delta}}{m_\pi} = \frac{1}{m_\pi a} + \frac{1}{2} m_\pi r\left(\frac{k^2}{m_\pi^2}\right)
                              + P  \left(\frac{k^2}{m_\pi^2}\right)^2 +\ldots
\label{eq:effrange}
\end{equation}
where $m_\pi a$, $m_\pi r$, and $P$ are called as the scattering length, effective range,
and shape parameter, respectively.
Note that $m_\pi a$ and $m_\pi r$ are cited in units of $m_\pi^{-1}$,
and an alterative way is employed in the present study
to denote the dimensionless quantity shape parameter $P$,
as compared with that of the isospin-$2$ $\pi\pi$ scattering in Ref.~\cite{Beane:2011sc},
where $P$ is scaled with $(m_\pi r)^3$.
Moreover, there is no minus sign in the first term in Eq.~(\ref{eq:effrange})
since the scattering length $m_\pi a$ is positive value
for the isospin-$0$ $\pi\pi$ scattering~\cite{Gasser:1983yg,Bijnens:1997vq,Beane:2011sc}.
For simple notations, $m a \equiv ma_0^\mathrm{I=0}$,
similar for $m_\pi r$ and $P$.

\begin{table}[t!]
\caption{Summaries of the effective range expansion parameters evaluated
from the lattice determinations of $k\cot\delta/m_\pi$ for three MILC lattice ensembles.
}
\label{tab:fitstoERT}
\begin{ruledtabular}
\begin{tabular}{c  c  c  c  }
$\rm Ensemble$ & Quantity  &   Fit A   &     Fit B    \\
\hline
\multirow{6}*{$(0.0036, 0.018)$}
&$m_\pi a$          & 1.72(25)     &  1.76(27) \\
&$m_\pi r$          & -0.818(689)  & -1.27(73)  \\
&$m_\pi^2 ar$       & -1.40(1.20)  & -2.25(1.32) \\
&$P$                & ---          & 0.788(727) \\
&$\chi^2/{\rm dof}$ & 0.542/2~\footnote{
Four data are within the region $k^2/m_\pi^2<0.5$ for this ensemble.
}
                                   &  0.327/2    \\
&$R_{a}$            & ---          & 12.5       \\
\hline
\multirow{6}*{$(0.0031, 0.031)$}
&$m_\pi a$          & 0.722(78)    &  0.736(79) \\
&$m_\pi r$          & -2.00(77)    & -2.24(65)    \\
&$m_\pi^2 ar$       & -1.44(58)    & -1.65(51)  \\
&$P$                & ---          &  0.592(648) \\
&$\chi^2/{\rm dof}$ & 0.37/1       &  1.08/2  \\
&$R_{a}$            & ---          &  4.73  \\
\hline
\multirow{6}*{$(0.0031, 0.0031)$}
&$m_\pi a$          & 0.730(97)    &  0.743(93) \\
&$m_\pi r$          & -1.69(1.02)  & -1.75(62)  \\
&$m_\pi^2 ar$       & -1.23(76)    & -1.30(49)  \\
&$P$                & ---          &  0.456(645) \\
&$\chi^2/{\rm dof}$ & 0.59/1       &  0.923/2   \\
&$R_{a}$            & ---          &  2.67  \\
\end{tabular}
\end{ruledtabular}
\end{table}

As it is pointed out in Ref.~\cite{Beane:2011sc},
the afore-mentioned  effective range approximation is believed to be convergent
for the energies below the $t$-channel cut,
which starts at ${k^2}={m_\pi^2}$.

\begin{figure}[t!]
\includegraphics[width=8.0cm,clip]{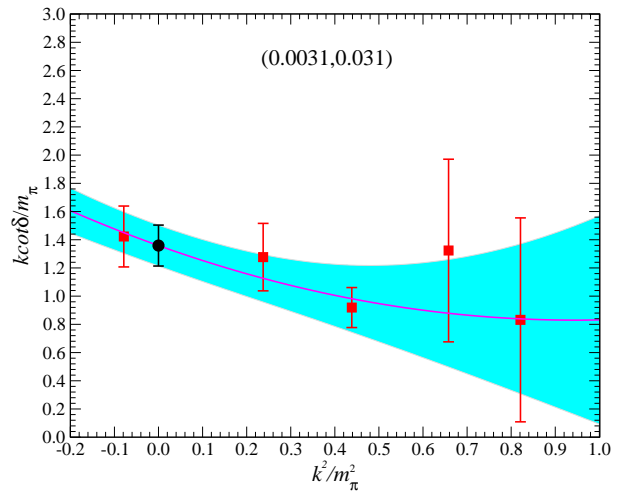}
\caption{
A three-parameter fit to lattice-determined values of $k\cot\delta/m_\pi$ over the region $k^2/m_\pi^2<1.0$ (Fit B)
for the $(0.0031,0.031)$  ensemble.
The shaded cyan band corresponds to statistical error, and
solid magenta curve is the central values.
The black circle in this figure indicates the relevant fit value of $1/(m_\pi a)$.
}
\label{fig:pipiEREfitA}
\end{figure}

\begin{figure}[!t]
\includegraphics[width=8.0cm,clip]{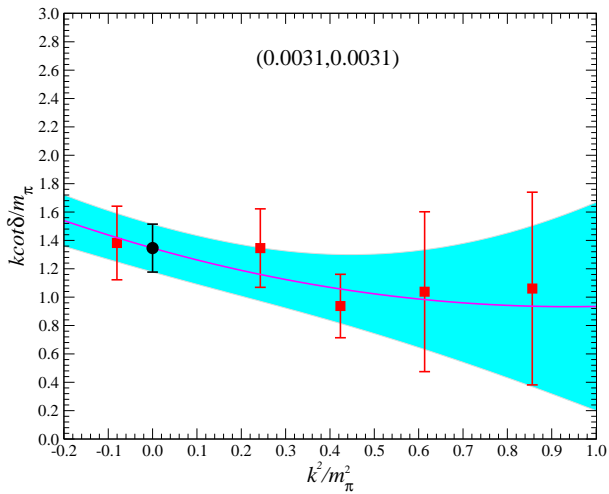}
\caption{
The same of Fig.~\ref{fig:pipiEREfitA}, but for the $(0.0031,0.0031)$  ensemble.
}
\label{fig:pipiEREfitB}
\end{figure}

\begin{figure}[!t]
\centering
\includegraphics[width=8.0cm,clip]{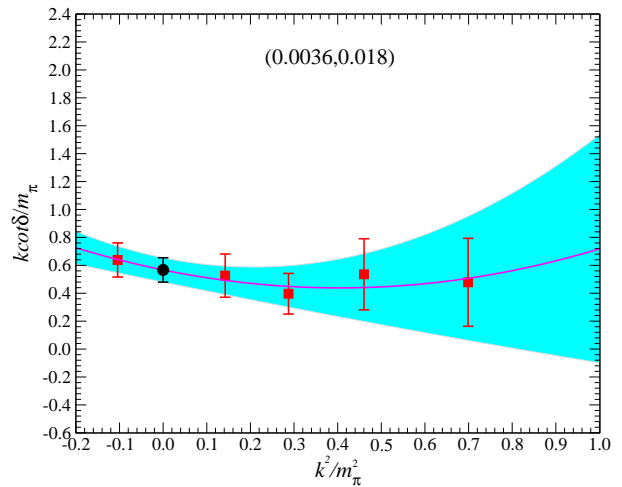}
\caption{
The same of Fig.~\ref{fig:pipiEREfitA}, but for the $(0.0036,0.018)$  ensemble.
}
\label{fig:pipiEREfitC}
\end{figure}

In the  present study, we just have five lattice data for each lattice ensemble at disposal.
This is mainly due to our currently limited computational resources.
Of course,  the lack of lattice ensembles with the different spatial extent $L$
for a given pion mass is an another important reason.
The lattice-determined values of $k \cot{\delta}/m_\pi$
are summarized in Table~\ref{tab:pp_Io_kcotk},
we also exhibit these values in Figs.~\ref{fig:pipiEREfitA},~\ref{fig:pipiEREfitB},~\ref{fig:pipiEREfitC}
for the $(0.0031, 0.031)$, $(0.0031, 0.0031)$ and $(0.0036, 0.018)$ ensembles, respectively.
Fortunately, the lattice-determined values of $k \cot{\delta}/m_\pi$
are turned out to be all within the $t$-channel cut ${k^2}={m_\pi^2}$.

Moreover, we observe that the values of $k\cot\delta/m_\pi$ are
not roughly linear in $k^2$ during the region $k^2/m_\pi^2<1.0$,
which reflects the fact that the shape parameter $P$
indeed has a impact on the curvature.
Actually, according to our analytical discussions in Sec.~\ref{sec:chiextrap},
the second term and third term in Eq.~(\ref{eq:effrange})
both contribute significantly for the values of $k\cot\delta/m_\pi$,~\footnote{
According to the discussions in Sec.~\ref{sec:chiextrap},
we can estimate that the ratio of the effective range $m_\pi r$ to
the shape parameter $P$ at the physical point is $-2.27(39)$,
which partially confirmed the assumption~\cite{Liu:2016cba}
that the contribution of ${\cal O}(k^4)$ term is not big than
that of ${\cal O}(k^2)$ term at least within $t$-channel cut ${k^2}={m_\pi^2}$.
}
Consequently, to be safe, we will include both terms
in the analyses of the values
of $k \cot{\delta}/m_\pi$ throughout the remaining analyses.

So far,  the influences of the higher order terms (NNLO, {\it etc})
from $\chi$PT is quite limited, which indicates the contributions from  ${\cal O}(k^6)$  term or higher are not clear~\cite{Liu:2016cba}.
Additionally, using just five lattice data for each  ensemble,
we are not able to effectively consider such high order contribution.

In the region $k^2/m_\pi^2<0.5$, we only have three or four data at disposal.
As is done for the isospin-$2$ $\pi\pi$ scattering in Ref.~\cite{Beane:2011sc},
the scattering length $m_\pi a$ and effective range $m_\pi r$
are fit (Fit A) with Eq.~(\ref{eq:effrange}), where the shape parameter $P$
and the other higher orders terms are fixed to be zero.
The fitted values of $m_\pi a$ and $m_\pi r$  are given in the third Column of Table~\ref{tab:fitstoERT}.

In the region $k^2/m_\pi^2<1$, lattice calculations indicates that
the curvatures have the quadratic (and higher) dependence on $k^2$.
Therefore, three leading effective range expansion parameters in Eq.~(\ref{eq:effrange})
are fit (Fit B) to lattice evaluations of $k\cot\delta/m_\pi$.
The fits are compared to the lattice calculations in Figs.~\ref{fig:pipiEREfitA},~\ref{fig:pipiEREfitB},~\ref{fig:pipiEREfitC}
for the $(0.0031, 0.031)$, $(0.0031, 0.0031)$ and $(0.0036, 0.018)$ ensembles, respectively.
The fitted values of $m_\pi a$, $m_\pi r$ and $P$ are given in the fourth Column of Table~\ref{tab:fitstoERT}.
The shaded cyan band corresponds to statistical error,
and the solid magenta curves are the central values.
The black circles in these figures indicate the relevant fit values of $1/(m_\pi a)$.

It is explicit from Table~\ref{tab:fitstoERT} that
the fit parameters of Fit B are reasonable consistent with those of Fit A
within the statistical  uncertainties.
In what follows, we will make use of $\chi$PT to predict the scattering parameters at physical pion mass,
with the fit parameters listed in Table~\ref{tab:fitstoERT} as input.
We note that our lattice-measured values of $k\cot\delta/m_\pi$ for each lattice ensemble
have relatively large errors, which give rise to the large statistical errors for the extracted quantities ($m_\pi a$, $m_\pi r$ and $P$).
On the same time, it leads to the relatively low $\chi^2/{\rm dof}$ values.
The straightforward way to move the $\chi^2/{\rm dof}$ into
a more reasonable range is to use more gauge configurations
for each lattice ensemble~\cite{Lepage:1989hd,Fukugita:1994ve,Fu:2016itp},
which is certainly beyond the scope of this work
since it requires the huge of the computer resource.
We note that some other fitting strategies to improve the statistical errors
are discussed in Refs.~\cite{Hu:2017wli,Hu:2016shf},

Usually, the truncation of the effective range $r$ is considered to be an important source of
systematic error, and in Ref.~\cite{Fu:2013ffa}, we only use one point (center of mass $\mathbf{P}=[0,0,0]$) to
evaluate the scattering length $(m_\pi a)_{000}$,
which is also calculated using only one data from center of mass frame
in the pioneering works~\cite{Kuramashi:1993ka,Fukugita:1994ve}.
To make this difference of these results more intuitive,
we also listed the relative error,
$$R_{a} = \left|\frac{m_\pi a-(m_\pi a)_{000}}{m_\pi a}\right|\times100\%\,,$$
in Table~\ref{tab:fitstoERT} for three lattice ensembles.
From Table~\ref{tab:fitstoERT}, we find that the relative errors
for three lattice ensembles are not small,
which means that this kind of the systematic error can not be ignored.
We view this as one of the most important results in the present study.

Considering the relatively strong interaction in
the isospin-$0$ $\pi\pi$ channel,
Liu {\it et al.} realized that the contribution of
$\mathcal{O}(k^2)$ and higher order terms in the
effective range expansion is very important for the reliable lattice calculation
of scattering length~\cite{Liu:2016cba}.
Since they have one energy level for each pion mass,
the values of the effective range $r$ are directly determined
from NLO $\chi$PT~\cite{Liu:2016cba}.
Hence, our investigation in this work
exactly follows Liu {\it et al.}'s work~\cite{Liu:2016cba},
and makes an improvement towards a calculation of $I=0$ $\pi\pi$ scattering length
by directly estimating the effective range $r$ from lattice simulation.

In addition,  we found that the shape parameter $P$
should be also included for the successful fit of  the low momentum lattice values of $k\cot\delta/m_\pi$ during the region $k^2/m_\pi^2<1$,
which are illustrated in Figs.~\ref{fig:pipiEREfitA},~\ref{fig:pipiEREfitB},~\ref{fig:pipiEREfitC}, respectively.
This will be clearly interpreted by the chiral perturbation prediction at NLO in Sec.~\ref{sec:chiextrap}.

Admittedly, our fitted values of the shape parameter $P$ listed in Table~\ref{tab:fitstoERT}
contain the rather large statistical errors, as a result,
it is not convincing to use these data to
perform the chiral extrapolation to the physical pion mass.
This kinds of work should be waiting for the more robust lattice data in the future.
Of course, the most efficient way to improve the statistical errors of the shape parameter $P$ is
working on the lattice ensembles with the different size $L$ for a given pion mass,
as is done for the isospin-$2$ $\pi\pi$ scattering in Ref.~\cite{Beane:2011sc},
since it can produce more reliable data in the region $k^2/m_\pi^2<1.0$.

Furthermore, according to the analytical arguments in Refs.~\cite{Lepage:1989hd,Fukugita:1994ve}
and rule of thumb discussions in Ref.~\cite{Fu:2016itp},
most straightforward way to improve the noise-to-signal ratios of the relevant correlators is
to employ very fine gauge configurations or anisotropic gauge configurations~\cite{Briceno:2016mjc}.
In addition, if we use lattice ensembles with relatively large  $L$,
and sum $\pi\pi$ correlators over all  time slices,
the signals are anticipated to be significantly improved~\cite{Fu:2016itp}.
With this aim in mind, our ongoing lattice investigation on the isospin-$0$ $\pi\pi$ scattering
will be carried out with MILC ultrafine
gauge configuration ($a \approx0.45$~fm and $L^3 \times T = 64^3\times 192$).

About five years ago, when we sincerely request Professor Eulogio Oset to
give us some comments for our previous work in Ref.~\cite{Fu:2013ffa},
he suggests us to consider the effective range parameter $r$ in the calculation of the $I=0$ $\pi\pi$ scattering length~\footnote{
The numerical computations of this work were unceasingly carried out for more than five years.
We should especially thank Prof. Hou Qing, and Prof. He Yan's continuous encouragements and
comprehensive supports.
}.
This work is specially to answer his inquiries, and we here admire
his sharp insight of the physical essence.

\newpage
\section{Chiral  extrapolations}
\label{sec:chiextrap}
The low-energy theorems imposed by chiral symmetry indicates that
each scattering parameter can be related to the relevant LEC
which appears at NLO in $\chi$PT~\cite{Beane:2011sc}.
In this section, we will exploit NPLQCD Collaboration's technique in Sec.~V of Ref.~\cite{Beane:2011sc}£¬
and follow the original derivations and notations in Refs.~\cite{Gasser:1983yg,Bijnens:1997vq,Beane:2011sc}
to provide the NLO $\chi$PT expressions for the effective range expansion and threshold parameters in the isospin-$0$ $\pi\pi$ scattering.
For easy notation, the Mandelstam variables ($s,u,t$) are expressed
in units of the physical pion mass squared $m_\pi^2$.

\subsection{Threshold parameters in $\chi$PT}
\label{sec:TPCHIPT}
In the elastic region ($4 \le s \le 16$),
unitarity indicates that the isospin-$0$  $\pi\pi$ $s$-wave scattering amplitude
$t(s) \equiv  t_{\ell=0}^{I=0}(s)$ can be described by real phase shift $\delta$~\cite{Gasser:1983yg,Bijnens:1997vq}
\begin{eqnarray}
t(s)= \Big(\frac{s}{s-4}\Big)^{\frac{1}{2}}\frac{1}{2i} \left\{e^{2i\delta(s)}-1\right\}\ .
\label{eq:unitary}
\end{eqnarray}
According to the discussions in Appendix~\ref{app:ChPT NLO},
the $I=0$ $\pi\pi$ $s$-wave NLO scattering amplitude can then be expressed in terms of
only three independent low-energy constants: $C_1$, $C_2$, and $C_3$~\cite{Gasser:1983yg,Bijnens:1997vq}:
\begin{widetext}
\begin{eqnarray}
\label{eq:pipinlo}
t(k)  &=& \frac{7m_\pi^2}{16\pi f_\pi^2} +
          \frac{m_\pi^4}{f_\pi^4}\left(C_1 + \frac{5}{384\pi^3} \right) +
          \frac{k^2}{f_\pi^2}\left[ \frac{1}{2\pi} +
          \frac{m_\pi^2}{f_\pi^2}\left(C_2 + \frac{247}{576\pi^3} \right) \right] +
          \frac{k^4}{f_\pi^4} \left(C_3 + \frac{35}{48\pi^3} \right) \cr
     && -\frac{1}{4\pi^3f_\pi^4}\left(\frac{63}{64}m_\pi^4 +  \frac{13}{6} m_\pi^2 k^2 + \frac{25}{18}k^4\right)\ln\left(\frac{m_\pi^2}{f_\pi^2}\right) \cr
     && + \frac{1}{16\pi^3f_\pi^4}\left(\frac{49}{16}m_\pi^4 +  7m_\pi^2 k^2 + 4k^4\right)\sqrt{\frac{k^2}{k^2+m_\pi^2}}
\ln\left(\frac{\sqrt{\frac{k^2}{k^2+m_\pi^2}}-1}{\sqrt{\frac{k^2}{k^2+m_\pi^2}}+1}\right) \cr
     && + \frac{1}{16\pi^3f_\pi^4}\left(\frac{5}{48}m_\pi^4 +  \frac{1}{3} m_\pi^2 k^2 + \frac{7}{9}k^4\right)\sqrt{\frac{k^2+m_\pi^2}{k^2}}
\ln\left(\frac{\sqrt{\frac{k^2+m_\pi^2}{k^2}}-1}{\sqrt{\frac{k^2+m_\pi^2}{k^2}}+1}\right)
\end{eqnarray}
\end{widetext}
where, $k=|{\bf k}|$ is the magnitude of  the center-of-mass three-momentum of each pion,
{\it i.e.}, $s=4(1+k^2/m_\pi^2)$.
and the constants $C_i(i=1,2,3)$ can be described in terms of the renormalized, quark mass independent low-energy constants  $\ell_i^r(\mu=f_{\pi,\mathrm{phy}})$~\cite{Gasser:1983yg,Bijnens:1997vq}
\begin{eqnarray}
\label{eq:c123_r}
C_1 &=& \frac{1}{4\pi}(20\ell_1^r + 20\ell_2^r + 5\ell_3^r + 7 \ell_4^r) + \frac{35}{256 \pi^3}   \cr
C_2 &=& \frac{1}{\pi}(16\ell_1^r + 12\ell_2^r + 2\ell_4^r) -  \frac{13}{192\pi^3}   \cr
C_3 &=& \frac{1}{3\pi}(44 \ell_1^r + 28 \ell_2^r )  - \frac{295}{576\pi^3}  .
\end{eqnarray}
It should be worthwhile to stress that three constants:
$5/(384\pi^3)$, $247/(576\pi^3)$, and $35/(48\pi^3)$ in Eq.~(\ref{eq:pipinlo})
will be neatly cancelled out in Taylor expansion of
last two terms in Eq.~(\ref{eq:pipinlo}) for small $k^2$.
Note that the renormalization scale $\mu$ is
fixed to be the physical pion decay constant $f_{\pi,\mathrm{phy}}$.
To express the relevant formulae as a function of $m_\pi/f_\pi$,
we change $f_{\pi,\mathrm{phy}}$ with $f_\pi$,
which only results in the corrections at NNLO~\cite{Liu:2016cba}.

The near threshold (to be specific, $k^2\rightarrow 0$) behavior for
the real part of the partial wave amplitude $t(k)$
can be expressed as a power-series expansion in the center-of-mass three-momentum~\cite{Gasser:1983yg,Bijnens:1997vq}
\begin{eqnarray}
\mbox{Re}\; t(k)\ =\ m_\pi a +k^2 b + k^4 c + {\cal O}(k^6),
\label{eq:threshexp}
\end{eqnarray}
where the threshold parameters $a$ and $b$ are referred to as the scattering length and  slope parameter, respectively.
Of course, the threshold parameter $c$ is also regarded as an another slope parameter.
Note that there is no minus sign before $m_\pi a$ in Eq.~(\ref{eq:threshexp}), which is consistent with the definition of the effective range approximation in Eq.~(\ref{eq:effrange}).

We should remark at this point that three independent constants $C_1, C_2$ and $C_3$
are directly related to the scattering length $a$, slope parameter $b$,
and slope parameter $c$, respectively.
This is a little bit different from those for the isospin-$2$ $\pi\pi$ scattering in Ref.~\cite{Beane:2011sc},
where three independent constants $C_1, C_2$ and $C_4$ are directly related to the scattering length $a$,
effective range $r$, and  shape parameter $P$, respectively.

It is easy to show that the inverse form of real part of scattering amplitude $t(k)$ in Eq.~(\ref{eq:unitary}) can be written as an elegant form,
\begin{equation}
[\mbox{Re}\, t(k)]^{-1} \hspace{-0.1cm}=\hspace{-0.1cm} \left(1+\frac{k^2}{m_\pi^2}\right)^{\hspace{-0.1cm}-1/2} 
\hspace{-0.1cm}\left[\frac{k\cot{\delta}}{m_\pi} + 
\frac{k^2}{m_\pi^2}\hspace{-0.1cm} \left({\frac{k\cot{\delta}}{m_\pi}}\right)^{\hspace{-0.1cm}-1}\right]\hspace{-0.1cm} .
\label{eq:threshexp2}
\end{equation}

Plugging Eq.~(\ref{eq:effrange}) into Eq.~(\ref{eq:threshexp2}),
and comparing them with the near threshold behavior of the partial  wave amplitude $t(k)$
denoted in Eq.~(\ref{eq:threshexp}), the effective range $m_\pi r$ and
the shape parameter $P$ can be nicely described
only in terms of three threshold parameters~\cite{Bijnens:1997vq}:
\begin{widetext}
\begin{equation}
m_\pi r  = \frac{1}{m_\pi a}-\frac{2 m^2_\pi b}{(m_\pi a)^2} - 2 m_\pi a \ ;
\label{eq:m_pi_r}
\end{equation}
\begin{equation}
P = \frac{8 m_\pi^4\left(b^2 - m_\pi a c\right) - 4m_\pi^2 b\left[m_\pi a + 2(m_\pi a)^3\right]
    -(m_\pi a)^2 + 4 (m_\pi a)^4  - 8(m_\pi a)^6} {8(m_\pi a)^3}  .
\label{eq:m_pi_P}
\end{equation}
\end{widetext}
Equations~(\ref{eq:m_pi_r}) and (\ref{eq:m_pi_P}) can be used inversely
to secure the slope parameters $b$ and $c$
from the lattice-measured effective range expansion parameters~\cite{Beane:2011sc}.

To simplify the notation, it is convenient to follow the original notation
in Ref.~\cite{Beane:2011sc} to denote $z\equiv m_\pi^2/f_\pi^2$.
After expanding the NLO partial wave scattering amplitude
denoted in Eq.~(\ref{eq:pipinlo}) in powers of $k^2$,
it is straightforward  to obtain the NLO $\chi$PT expressions for the threshold parameters:
\begin{eqnarray}
m_{\pi} a   &=&  \frac{7z}{16\pi}   +    z^2 C_1 - \frac{63}{256\pi^3} z^2 \log z , \label{eq:Thresholds_a} \\
m_\pi^2 b   &=&  \frac{z}{2\pi}     +    z^2 C_2 - \frac{13}{ 24\pi^3} z^2 \log z , \label{eq:Thresholds_b} \\
m_\pi^4 c   &=&  z^2 C_3 - \frac{25}{72\pi^3} z^2\log z .
\label{eq:Thresholds_eforms}
\end{eqnarray}
It is interesting and important to notice that no any contributions
are accepted from LO $\chi$PT for the slope parameter $c$,
as it is already noticed that for the isospin-$2$ $\pi\pi$ scattering in Ref.~\cite{Beane:2011sc}.

In our previous work~\cite{Fu:2013ffa}, we directly make use of
the explicit results in Appendix C of Ref.~\cite{Bijnens:1997vq}
to get the $\chi$PT formula for the scattering length $m_\pi a$,
we find it is nice to consistent with the relevant result in Eq.~(\ref{eq:Thresholds_a}),
which was recently  used to extrapolate the lattice-measured data to
the physical pion mass by Liu {\it et al.} in Ref.~\cite{Liu:2016cba}.
Note that the constant $C_1$ is related to the constant $\ell_{\pi\pi}^{I=0}$ denoted in  Ref.~\cite{Fu:2013ffa} by
\begin{equation}
C_1 =  \frac{7}{256\pi^3} (\ell_{\pi\pi}^{I=0}+5) .
\label{eq:C1_2_pp}
\end{equation}
As it demonstrated in Appendix~\ref{app:ChPT NLO},
we also nicely reproduce the relevant results in Appendix C of Ref.~\cite{Bijnens:1997vq}
for slope parameter $b$.
Note that there is no explicit formula for the slope parameter $c$ in Ref.~\cite{Bijnens:1997vq}.

From Eq.~(\ref{eq:m_pi_r}), we note that the effective range $m_\pi r$ contains three components.
It is easy to show that the second term indeed dominates the effective range $m_\pi r$ for
the small  $z$-values, as it is observed in Ref.\cite{Liu:2016cba},
on the other hand, for the large $z$-values,
the third term absolutely dominates the effective range $m_\pi r$,
and the first term approaches zero.
To be safe, we include all the three parts in the present study.

Plugging these formulas for the threshold parameters
[namely,  the equations~(\ref{eq:Thresholds_a}), (\ref{eq:Thresholds_b}) and
 (\ref{eq:Thresholds_eforms})] into the equations~(\ref{eq:m_pi_r}) and (\ref{eq:m_pi_P}),
 after some strenuous algebraic operations,
it is now straightforward to achieve the NLO $\chi$PT descriptions for the effective range approximation parameters:
\begin{eqnarray}
\hspace{-0.5cm}m_\pi r &=& -\frac{144\pi}{49z} + C_4   + \frac{157}{147\pi} \ln z
- \frac{7z}{8\pi} - 2 z^2 C_1 , \label{eq:mpr_NLO} \\
\hspace{-0.5cm}m_\pi^2 a r             &=& -\frac{9}{7} + z C_5 + \frac{25}{21\pi^2} z\ln z
-\frac{49z^2}{128\pi^2} -  \frac{7z^3}{4\pi} C_1, \label{eq:mpi_ar_NLO} \\
\hspace{-0.5cm}P       &=& \frac{478\pi}{343z} + C_6 + \frac{673}{3528\pi} \ln z
- \frac{9z}{32\pi} \cr
&+& \frac{8192\pi^3}{343}C_3 C_1 z \label{eq:mpiP_NLO}
 + \left(\frac{10240}{1029} C_1 - \frac{288}{49}C_3 \right) z\ln z \cr
&+& z^2  \left(\frac{1}{2}C_1-C_2\right)
- \frac{343z^3}{4096\pi^3} - \frac{147 z^4}{256\pi^4} C_1,
\label{eq:extrapforms}
\end{eqnarray}
where the constants $C_4$, $C_5$ and $C_6$  are solely denoted
in terms of the constants $C_1, C_2$, and $C_3$ via
\begin{eqnarray}
C_4 &=& \frac{256\pi^2}{49}\left(\frac{25}{7} C_1 - 2 C_2 \right) , \label{eq:extrapforms_C4} \\
C_5 &=& \frac{16\pi}{7} \left(\frac{16}{7} C_1 - 2 C_2 \right) , \label{eq:extrapforms_C5} \\
C_6 &=& -\frac{33248\pi^2}{2401}C_1 + \frac{3200\pi^2}{343}C_2 -\frac{256\pi^2}{49}C_3  .
\label{eq:extrapforms_C6}
\end{eqnarray}
Note that we should keep last two terms in NLO $\chi$PT expressions
for $m_\pi r$ in Eq.~(\ref{eq:mpr_NLO}),
which naturally come from the third part of $m_\pi r$ denoted in Eq.~(\ref{eq:m_pi_r}).
In the same manner, the last two terms in NLO $\chi$PT expressions for $m_\pi^2 ar$ in Eq.~(\ref{eq:mpi_ar_NLO})
are required to hold as well.
On the other hand, the relevant two terms for the isospin-$2$
case can be reasonably overlooked~\cite{Beane:2011sc}.
Similarly, the last six terms in NLO $\chi$PT expressions
for $P$ in Eq.~(\ref{eq:mpiP_NLO}) should be reserved.

We should remark at this point that Eqs.~(\ref{eq:mpr_NLO}), (\ref{eq:mpi_ar_NLO}),
and (\ref{eq:mpiP_NLO}) are valid
for the range of interest in the present study,
and more terms should be added into these equations for large $z$-values.
Moreover, the factor $-9/7$  in Eq.~(\ref{eq:mpi_ar_NLO}) indicates the relevant
LO $\chi$PT prediction.

In practice, it is handful to recast the constants $C_i (i=1,2,3)$ in terms of
the scale-independent dimensionless couplings
$\overline{\ell}_i$~\cite{Gasser:1983yg,Bijnens:1997vq}
\begin{eqnarray}
C_1 &=&  \hspace{-0.02cm}\frac{1}{256\pi^3}\hspace{-0.1cm}\left[\hspace{-0.1cm}\frac{40}{3} \bar{\ell}_1
\hspace{-0.05cm}+\hspace{-0.05cm} \frac{80}{3}\bar{\ell}_2 \hspace{-0.05cm}-\hspace{-0.05cm} 5\bar{\ell}_3 \hspace{-0.05cm}+\hspace{-0.05cm} 28 \bar{\ell}_4
\hspace{-0.05cm}+\hspace{-0.05cm} 63 \ln\frac{m_\pi^2}{f_{\pi,\mathrm{phy}}^2} \hspace{-0.05cm}+\hspace{-0.05cm} 35\right],   \cr
C_2 &=& \frac{1}{192\pi^3}\left[ 32 \bar{\ell}_1 +  48 \bar{\ell}_2  +  24 \bar{\ell}_4
+ 104 \ln\frac{m_\pi^2}{f_{\pi,\mathrm{phy}}^2} - 13\right],     \cr
C_3 &=&  \frac{1}{72 \pi^3}\left[11 \bar{\ell}_1 + 14 \bar{\ell}_2
    + 25\ln\frac{m_\pi^2}{f_{\pi,\mathrm{phy}}^2}  - \frac{295}{8} \right] .
\label{eq:C123b}
\end{eqnarray}

Meanwhile, using Eq.~(\ref{eq:extrapforms_C4}) and Eq.~(\ref{eq:extrapforms_C5}),
we can also recast the $C_4$ and $C_5$ in terms of
the scale-independent dimensionless couplings
\begin{eqnarray}
C_4 &=& \frac{1}{147\pi}\left[
-\frac{792}{7} \bar{\ell}_1 - \frac{688}{7}  \bar{\ell}_2
-\frac{375}{7}  \bar{\ell}_3 + 108  \bar{\ell}_4 \right. \cr
&&\left.
- 157\ln\frac{m_\pi^2}{f_{\pi,\mathrm{phy}}^2} + 479 \right] , \cr
C_5 &=&
-\frac{1}{42\pi^2} \left[
\frac{144}{7} \bar{\ell}_1 +  \frac{176}{7}  \bar{\ell}_2
+ \frac{30}{7}  \bar{\ell}_3 \right. \cr
&&\left.
+ 50\ln\frac{m_\pi^2}{f_{\pi,\mathrm{phy}}^2}
+ 43 \right] .
\label{eq:C45_b}
\end{eqnarray}

Using Eq.~(\ref{eq:C123b}) and Eq.~(\ref{eq:C45_b}), we can measure the values of $C_1$, $C_2$, $C_3$, $C_4$, and $C_5$ via the reported values of the scale independent couplings $\bar{\ell}_i$
in Refs.~\cite{Colangelo:2001df,Bijnens:2014lea}
\begin{eqnarray}
\bar{\ell}_1 &=& -0.4\pm0.6\,, \qquad \bar{\ell}_2 = 4.3\pm0.1 \,, \cr
\bar{\ell}_3 &=& 2.9\pm2.4\,,\,\,\,\,\, \qquad \bar{\ell}_4 = 4.4\pm0.2 \,,
\end{eqnarray}
and the renormalization scale $\mu$ is set to be the physical pion decay constant
$f_{\pi,\mathrm{phy}}=130.2(1.7)$~\cite{Patrignani:2016xqp}.
Then we arrive at
\begin{eqnarray}
C_1 &=& 0.033(2) \,, \quad C_2 = 0.051(4) \,, \quad C_3 = 0.010(3), \cr
C_4 &=& 0.865(0.511) \,, \quad C_5 = -0.184(59)   .
\label{eq:C123_CGL}
\end{eqnarray}
Note that if we make use of  Eq.~(\ref{eq:c123_r}) and take the estimated values of $\ell_i^r$'s in Ref.~\cite{Liu:2016cba},
we can get the same values of  $C_i (i=1,2,3,4,5)$.

With the estimated values of $C_1$, $C_2$, and $C_3$ in Eq.~(\ref{eq:C123_CGL}),
and using equations~(\ref{eq:m_pi_r}) and (\ref{eq:m_pi_P}),
the ratio of the effective range $m_\pi r$ to
the shape parameter $P$ at the physical pion mass can be estimated as $-2.27(39)$.
This indicates the second term and third term in Eq.~(\ref{eq:effrange})
both contribute significantly for the lattice-measured values of $k\cot\delta/m_\pi$.
Meanwhile, it partially confirmed the assumption in Ref.~\cite{Liu:2016cba}
that the contribution of the ${\cal O}(k^4)$ term is not big than
that of ${\cal O}(k^2)$ term at least within the $t$-channel cut ${k^2}={m_\pi^2}$.
Admittedly, at NLO $\chi$PT level, the contributions from  ${\cal O}(k^6)$  term or higher are not clear for us as well~\cite{Liu:2016cba},
and it definitely needs the knowledge of the higher order terms (NNLO, {\it etc}) from $\chi$PT,
which is certainly beyond the scope of this work.

\begin{figure}[!t]
\includegraphics[width=8.5cm,clip]{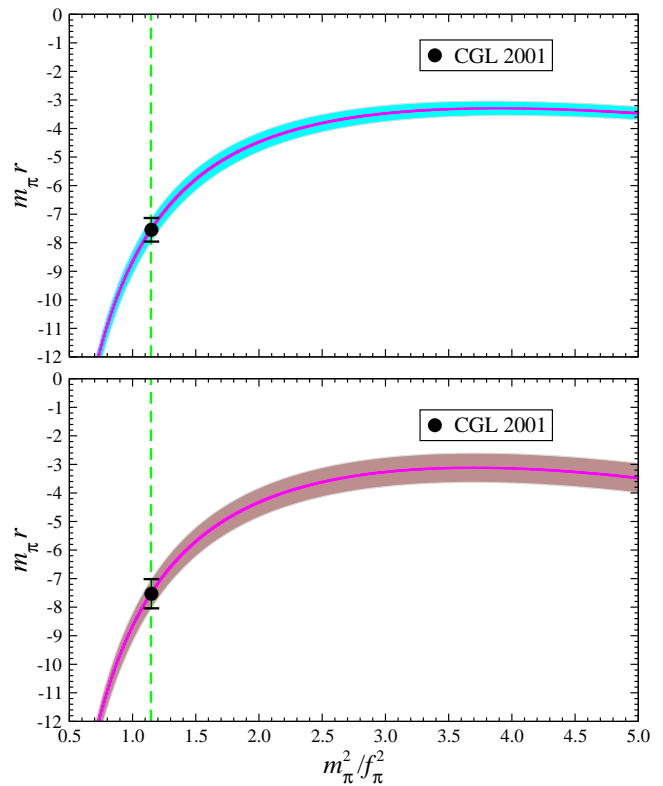}
\caption{
The dashed green lines denote the physical line,
and the Roy equation predictions~\protect\cite{Colangelo:2001df}
are the black circle on the physical line.
The brown band on bottom panel denotes the results are calculated at NLO in $\chi$PT using Eq.~(\ref{eq:mpr_NLO}) with the relevant input data from Ref.~\cite{Colangelo:2001df}.
For comparison, the cyan band on top panel denotes the relevant results are directly evaluated
using Eq.~(\ref{eq:m_pi_r}).
The solid magenta curves are the central values.
}
\label{fig:cgl_NLO_mr}
\end{figure}

It should be worthwhile to stress that, for the isospin-$2$ $\pi\pi$ scattering,
the $m_\pi r$ always has positive value~\cite{Beane:2011sc}.
However, for the isospin-$0$ case, it always holds negative value.
To verify the valid of the Eq.~(\ref{eq:mpr_NLO}),
we straightforwardly use Eq.~(\ref{eq:m_pi_r}) to measure the corresponding results,
where the Eqs.~(\ref{eq:Thresholds_a}) and (\ref{eq:Thresholds_b}) are used, and
the values of $C_1$, $C_2$, and $C_3$ in Eq.~(\ref{eq:C123_CGL}) are plugged in.
The relevant results  are displayed in top panel of Fig.~\ref{fig:cgl_NLO_mr}.
We also show the relevant results using Eq.~(\ref{eq:mpr_NLO}),
which are displayed in bottom panel of Fig.~\ref{fig:cgl_NLO_mr}.
One thing greatly comforting us is that both methods give similar results,
which indicates that Eq.~(\ref{eq:mpr_NLO}) is valid
at least for the range of interest in this work.
Analogously, we can discuss the validity of  Eq.~(\ref{eq:mpi_ar_NLO}).
In what follows, we can use Eq.~(\ref{eq:mpr_NLO}) and Eq.~(\ref{eq:mpi_ar_NLO}).

Similarly, to check the applicable scope of Eq.~(\ref{eq:mpiP_NLO}),
we directly make use of Eq.~(\ref{eq:m_pi_P}) to measure the corresponding results.
It is very inspiring that both methods also result in the somewhat  similar results,
which indicates that Eq.~(\ref{eq:mpiP_NLO}) is also valid
at least for the range of interest in the present study.
Admittedly, Eq.~(\ref{eq:mpiP_NLO}) is not simple style.
For the large $z$-values (for example, $z>6$), it deviates seriously from these using Eq.~(\ref{eq:m_pi_P}).
In the rest of the analysis, we can use Eq.~(\ref{eq:mpiP_NLO}).

It is worth mentioning that the NLO expressions given in Eqs.~(\ref{eq:mpr_NLO}),
(\ref{eq:mpi_ar_NLO}), and (\ref{eq:mpiP_NLO}) are much
more complicate than those of the isospin-$2$~\cite{Beane:2011sc}.
Actually, it partially reflects the fact that the LO $\chi$PT
can nicely reproduce the $I=2$ $s$-wave $\pi\pi$ scattering length
with just a $0.5\%$ deviation as compared with the relevant experimental
and theoretical results~\cite{Weinberg:1966kf, Batley:2010zza, Pislak:2003sv}.
Nevertheless, in the isospin-$0$ channel,
the agreement between LO $\chi$PT and corresponding experiments and theoretical results
is departed by about 30\%~\cite{Weinberg:1966kf, Batley:2010zza, Pislak:2003sv}.

\newpage
\subsection{Chiral extrapolation of threshold parameters}

In this work, lattice calculations are performed
at pion masses:  $247\, \rm MeV$, $249\, \rm MeV$ and $314\, \rm MeV$, respectively,
according to the previous discussions,
the influence of the $\sigma$ meson can be reasonably ignored.
In principle, we can exploit all of our data to carry out
the relevant chiral extrapolation.
Nevertheless, as it is pointed out in Ref.~\cite{Liu:2016cba},
the pion mass values should be small enough
to make the NLO chiral expansion valid.
For this purpose, we carry out the chiral extrapolation
only using the data with two lower pion masses (247 MeV and 249 MeV).

\begin{figure}[!t]
\includegraphics[width=8.5cm,clip]{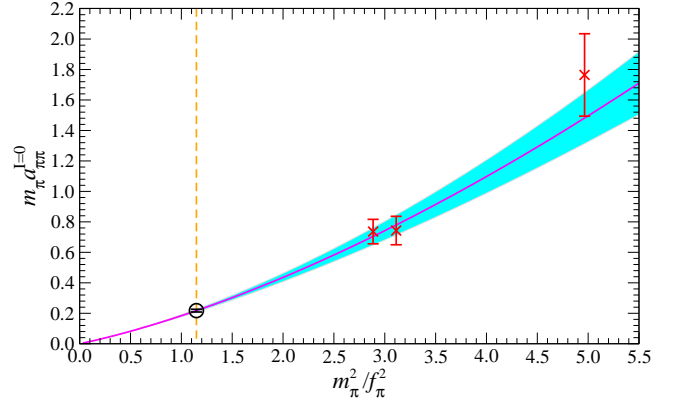}
\caption{
Chiral extrapolation of $m_\pi a $ using only the lattice data
with two lower pion masses.
The dashed orange line indicates the physical line.
The lattice QCD $+\chi$PT prediction at physical pion mass is
the black circle on the physical line,
where the statistical error and systematic  error are added in quadrature.
The shaded cyan band corresponds to statistical error, and
solid magenta curve is the central values.
}
\label{fig:maPHYS}
\end{figure}

Using  NLO $\chi$PT expressions for the scattering length $m_\pi a$
in Eq.~(\ref{eq:Thresholds_a}),
the value of the constant $C_1$ can be obtained
by fitting with lattice-measured $m_\pi a$
from Fit B provided in Table~\ref{tab:fitstoERT}.
We can then translate the constant $C_1$ into the familiar $\ell_{\pi\pi}^{I=0}$ by
Eq.~(\ref{eq:C1_2_pp}).
Moreover, the scattering length $m_\pi a$ at physical pion mass can be predicted
using NLO $\chi$PT expressions denoted in Eq.~(\ref{eq:Thresholds_a}).
The chiral extrapolation of the scattering length $m_\pi a $ is shown in Fig.~\ref{fig:maPHYS},
and the extrapolated value at physical point is indicated by black circle
on the physical line.
The relevant results are given in Table~\ref{tab:myFit_ar} as Fit-1.

On the same time, the corresponding fitting results
with all the tree pion masses are also provided
in Table~\ref{tab:myFit_ar} as Fit-2, which agree with Fit-1 within statistical errors.
The differences of two fits are considered
as the estimated  systematic uncertainties~\cite{Liu:2016cba}.
This leads to our ultimate results for scattering length $m_\pi a_0^\mathrm{I=0}$,  $\ell_{\pi\pi}^{I=0}$ and $C_1^{NLO}$  as
\begin{eqnarray}
\hspace{-1.0cm} m_\pi a_0^\mathrm{I=0} &=& 0.217(9)(5), \qquad \ell_{\pi\pi}^{I=0}    = 45.6(7.6)(3.8),\cr
\hspace{-1.0cm} C_1^{NLO}              &=& 0.0448(68)(31),
\label{eq:Csfit_a}
\end{eqnarray}
where the superscript in the constant $C_1$ indicates that
it is estimated at NLO $\chi$PT.
Our lattice result of $m_\pi a_0^\mathrm{I=0}$ is
in reasonable agreement with the newer experimental and theoretical
determinations as well as lattice calculations.
In Table~\ref{tab:CompMpia0}, we compare this result,
together with $\ell_{\pi\pi}^{I=0}$,
to these relevant results accessible in the literature.
\begin{table}[b!]
\caption{Results of NLO chiral fit for $m_\pi a$, $m_\pi r$ and $m_\pi^2 ar$.
Fit-1 uses two data from the $(0.0031,0.031)$  and $(0.0031,0.0031)$ ensembles,
while Fit-2 includes three data from all the  lattice ensembles.
}
\centering
\begin{tabular*}{\linewidth}{@{\extracolsep{\fill}}lcc}
\hline\hline
{\rm{Quantity}} & \rm{Fit-1} & \rm{Fit-2} \\
    \hline\hline
    $m_\pi a_0^\mathrm{I=0}$       & $0.217(9)$    & $0.222(8)$     \\
    $C_1^{NLO}$                    & $0.0448(68)$  & $0.0479(58)$   \\
    $\ell_{\pi\pi}^\mathrm{I=0}$   & $45.6(7.6)$   & $49.4(6.6)$    \\
    $\chi^2/\mathrm{dof}$          & $0.32/1$      & $1.1/2$        \\
    \hline
   $m_\pi r$                       & $-6.07(44)$     & $-5.71(38)$  \\
    $C_4^{NLO}$                    & $2.36(44)$      & $2.71(38)$   \\
    $\chi^2/\mathrm{dof}$          & $0.27/1$        & $2.42/2$     \\
     \hline
   $m_\pi^2 ar$                    & $-1.18(0.13)$   & $-1.13(0.12)$  \\
    $C_5^{NLO}$                    & $0.152(0.117)$  & $0.195(0.107)$ \\
    $\chi^2/\mathrm{dof}$          & $0.42/1$        & $1.27/2$       \\
\hline\hline
\end{tabular*}
\label{tab:myFit_ar}
\end{table}

\begin{table}[b!]
  \centering
  \caption{Comparison of results available in the literature for
  $m_\pi a$ and $\ell_{\pi\pi}^\mathrm{I=0}$.}
    \begin{tabular*}{\linewidth}{@{\extracolsep{\fill}}lrr}
    \hline\hline
    & $m_\pi a_0^\mathrm{I=0}$ & $\ell_{\pi\pi}^\mathrm{I=0}$ \\
    \hline\hline
    This work                  & $0.217(9)(5)$ & $45.6(7.6)(3.8)$          \\
    Liu~\cite{Liu:2016cba}     & $0.198(9)(6)$ & $30(8)(6)$       \\
    Fu~\cite{Fu:2013ffa}       & $0.214(4)(7)$ & $43.2(3.5)(5.6)$ \\

    \hline
    Weinberg~\cite{Weinberg:1966kf} & $0.1595(5)$ & $-$    \\
    CGL~\cite{Colangelo:2001df} & $0.220(5)$ & $48.5(4.3)$ \\

    \hline
    NA48/2~\cite{Batley:2010zza} & $0.220(3)(2)$ & $$     \\
    E865~\cite{Pislak:2003sv} & $0.216(13)(2)$ & $45.0(11.2)(3.5)$ \\
    \hline\hline
  \end{tabular*}
  \label{tab:CompMpia0}
\end{table}

With NLO $\chi$PT expressions for the effective range $m_\pi r$ in Eq.~(\ref{eq:mpr_NLO})
and $m_\pi^2 ar$ in Eq.~(\ref{eq:mpi_ar_NLO}),
the $C_4$ and $C_5$ values can be obtained by fitting with lattice-determined $m_\pi r$
and $m_\pi^2 ar$
from Fit B listed in Table~\ref{tab:fitstoERT},
where the $C_1$ is fixed to the cental value of $C_1^{NLO}$ in Eq.~(\ref{eq:Csfit_a}),
since we just have two lattice data at disposal.
Moreover, the effective range $m_\pi r$ and $m_\pi^2 ar$  at physical pion mass
can be predicted by NLO $\chi$PT.
The relevant results are provided in Table~\ref{tab:myFit_ar} as Fit-1.

\begin{figure}[!t]
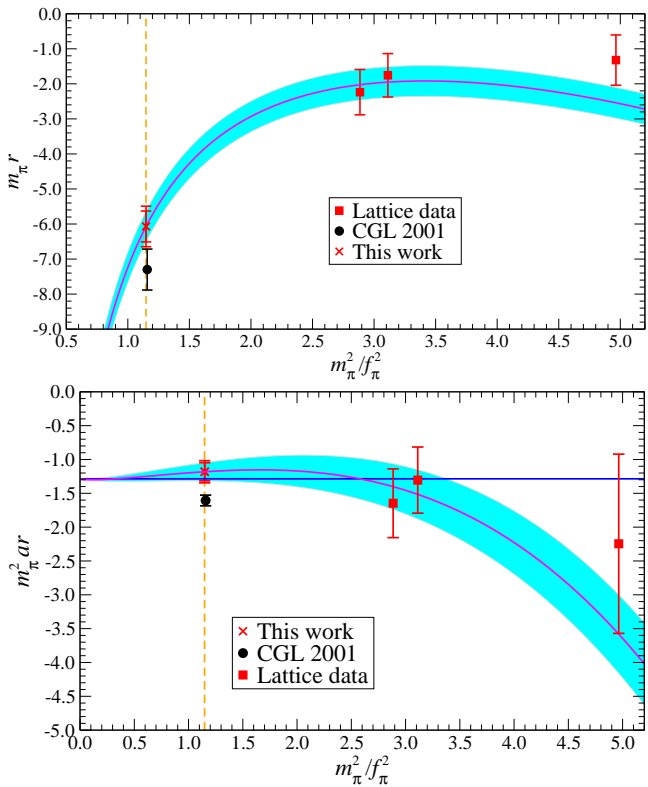

\includegraphics[width=8.5cm,clip]{mpi_r.eps}
\includegraphics[width=8.5cm,clip]{mpi_ra.eps}
\caption{
Chiral extrapolations of $m_\pi r $ and $m_\pi^2 ar $ using only the lattice data
with two lower pion masses.
The dashed orange line indicates the physical line.
The lattice QCD $+\chi$PT prediction at physical pion mass
is the red cross on the physical line,
where the statistical error and systematic  error are added in quadrature.
The Roy equation predictions~\protect\cite{Colangelo:2001df}
are the black circles on the physical line.
The horizontal blue solid line in the bottom panel denotes
the LO $\chi$PT prediction, which is $m_\pi^2 ar = -9/7 $  in the chiral limit.
}
\label{fig:mrPHYS}
\end{figure}

Besides, the relevant fitting results with three pion masses
are given in Table~\ref{tab:myFit_ar} as Fit-2,
which agree with Fit-1 within statistical errors.
The differences of two fits are regarded as the estimated  systematic errors.
This leads to our ultimate results of $m_\pi r$ and $m_\pi^2 ar$ as
\begin{equation}
m_\pi r   =  -6.07(44)(36), \quad m_\pi^2 ar   =  -1.18(13)(6).
\label{eq:Csfit_mr}
\end{equation}
In Fig.~\ref{fig:mrPHYS}, the chiral extrapolations of  $m_\pi r $ and $m_\pi^2 ar $
are shown in the top panel, and bottom panel, respectively.
Note that  our lattice determination of $m_\pi^2 ar$ is fairly consistent with
the LO $\chi$PT prediction of $m_\pi^2 ar = -9/7(1+{\cal{O}}(m_\pi^2/\Lambda_\chi^2))$.

It is quite inspiring that our lattice-obtained values of the scattering length $m_\pi a_0^\mathrm{I=0}$
the effective range $m_\pi r$, and $m_\pi^2 ar$  are turned out
to be in reasonable accordance with the Roy equation determinations~\cite{Colangelo:2001df},
\begin{eqnarray}
\hspace{-0.5cm} m_\pi a_0^\mathrm{I=0} &=& 0.220(5) ,   \qquad \hspace{0.5cm} m_\pi^2 b  = 0.276(6) , \cr
\hspace{-0.5cm} m_\pi r                &=& -7.300(0.584), \quad m_\pi^2 ar = -1.606(79).
\end{eqnarray}
Fig.~\ref{fig:mrPHYS} provides a comparison
of our lattice calculations and the Roy equation
values of $m_\pi r$ and $m_\pi^2 ar$, which are indicated by black circles on the physical line.
Our fit results of  $m_\pi a_0^\mathrm{I=0}$ and $m_\pi r$ seem to be reasonable consistent
with the corresponding CGL predictions roughly at the $1\sigma$-level~\cite{Colangelo:2001df}.
However, for $m_\pi^2 ar$, the roughly $2\sigma$-level discrepancy is observed.



According to Eq.~(\ref{eq:mpr_NLO}),
the effective range $m_\pi r$ is divergent in the chiral limit,
this can partially explain the extrapolated value of $m_\pi r$ to the physical point
usually has relative large statistical uncertainty,
as compared with those of $m_\pi a_{0}^{I=0}$ and  $m_\pi^2 ar$.
In practice, the statistical error of $m_\pi r$ is roughly estimated
by the error of $C_4$, and it is not dependent on the pion mass (or $z$-value).

\newpage
\section{Summary and Conclusion}
\label{sec:discussion}

In this work, using the MILC fine or superfine gauge configurations~\cite{Aubin:2004wf,Bernard:2001av}
with three flavors of Asqtad-improved staggered dynamical quarks~\cite{stag_fermion},
we performed a lattice study of the isospin-$0$ $s$-wave $\pi\pi$ scattering
over a range of momenta below the inelastic threshold, and total momenta
$\mathbf{P}=[0,0,0]$, $[0,0,1]$, $[0,1,1]$, $[1,1,1]$, and $[0,0,2]$,
where, L\"uscher's technique~\cite{Luscher:1986pf,Luscher:1990ux,Luscher:1990ck}
and its extensions\cite{Beane:2003da,Rummukainen:1995vs,Kim:2005gf,Christ:2005gi,Doring:2012eu,Fu:2011xz,Leskovec:2012gb,Doring:2012eu}
are utilized to extract the scattering phase shifts
with lattice-calculated energy-eigenstates.

The technique of the ``moving'' wall source introduced in Refs.~\cite{Kuramashi:1993ka,Fukugita:1994ve}
is exploited to calculate four diagrams classified
for the $I=0$ $\pi\pi$ scattering in Refs.~\cite{Kuramashi:1993ka,Fukugita:1994ve}.
Consequently, the signals of vacuum diagram are remarkably improved as compared with
our previous studies~\cite{Fu:2012gf,Fu:2013ffa},
which enables us to not only measure the scattering length,
but also explore the effective range.
The chiral extrapolations of $m_\pi a_{0}^{I=0}$ and $m_\pi r$  are performed by NLO $\chi$PT.
Extrapolated to the physical value of $m_\pi/f_\pi$,
our final outcomes yield
\begin{equation}
m_\pi a_0^\mathrm{I=0} = 0.217(9)(5) ,  \qquad
m_\pi   r              = -6.07(44)(36) , \nonumber
\end{equation}
which are in reasonable agreement with the newer experimental and theoretical
determinations as well as the lattice calculations.

In the interpolation of a fit to the lattice-measured values of $k\cot\delta/m_\pi$
during the region $k^2/m_\pi^2<1.0$,
we include three leading  parameters in the effective range expansion,
which implement Liu {\it et al.}'s strategies~\cite{Liu:2016cba} directly from lattice QCD.
In particular, we confirmed that the effective range $r$ and shape parameter $P$
should be included for the successful fit~\cite{Liu:2016cba}.

For each lattice ensemble, we just calculate five points,
simply due to the lack of computational resources,
thus, robust extraction of shape parameter $P$
definitely need more lattice data for each lattice ensemble.
Admittedly, the most efficient way to improve the statistical errors of $P$ is
working on lattice ensembles with different size $L$ for a given pion mass,
as  is done for the isospin-$2$ $\pi\pi$ scattering in Ref.~\cite{Beane:2011sc}.

So far,  the influences of the higher order terms
from $\chi$PT is quite limited,
and we also cannot rule out that such contributions are significant~\cite{Liu:2016cba}.
It will be very interesting to systematically study  the NNLO $\chi$PT expressions for leading three terms in the effective range expansion, and investigate ultimate numerical results
due to its modifications from the NLO $\chi$PT expressions,
we reserve this challenging work in the future.

The $\sigma$ meson is clearly presented in low energy~\cite{Sasaki:2010zz,Fu:2011wc},
and it is necessary to map out ``avoided level crossings''
between $\sigma$ resonances and isospin-$0$ $\pi\pi$ states
to secure the reliable scattering length as investigated in the $\pi K$ scattering~\cite{Sasaki:2010zz,Fu:2011wc}.
Luckily, according to Liu {\it et al.}'s discussions~\cite{Liu:2016cba},
the contaminations from $\sigma$ meson for three lattice ensembles
with small pion masses are negligible.
Of course, sigma meson should be incorporated into more sophisticated
lattice computation in the future.

\section*{Acknowledgments}
This work is partially supported by both the National Magnetic Confinement Fusion Program of China (Grant
No. 2013GB109000) and the Fundamental Research Funds for the Central Universities (Grant No. 2010SCU23002).
This project was partially supported by the High Performance Computing Center of Sichuan University, China.
We would like to deliver our sincere gratitude to MILC Collaboration
for permitting us to use the MILC gauge configurations and MILC codes.
We deeply thank  Carleton DeTar for his instructing Ziwen the necessary
theoretical knowledge and computational skills for this work.
We especially thank  Michael Doering for his enlightening and constructive comments.
We cordially express our boundless gratitude for the support of Professor Hou Qing
and Professor He Yan.
We would like to express our appreciation to Professor He Yan, Professor
Huang Ling, Professor Wang Jun, and Professor Fu Zhe, and other warm-hearted people
for donating us enough removable hard drives to store the propagators for the present work.
We also express gratitude to the Institute of Nuclear Science and Technology, Sichuan
University, from which the computer resources and electricity costs were furnished,
and numerical calculations were partially carried out at both PowerLeader
Clusters and AMAX, CENTOS, HP, and ThinkServer workstations.

\appendix
\section{FFT algorithm for vacuum diagram}
\label{app:FFT}
In this Appendix, we follow the original notations in Ref.~\cite{Bernard:2007qf}
to transplant the FFT algorithm for the disconnected piece for sigma operator~\cite{Bernard:2007qf}
into the vacuum diagram of $\pi\pi$ operator.

According to  Eq.~(\ref{eq:dcr}),  the first part of  the vacuum diagram
can be expressed with quark propagators $G$ as
\begin{eqnarray}
\label{app:dcr}
\hspace{-0.8cm} C_{\pi\pi}^V({\mathbf p},t_4,t_3,t_2,t_1) &=&
\mbox{Re} \sum_{\mathbf{x}_2,\mathbf{x}_3 }
e^{ i{\mathbf p} \cdot ({\mathbf{x}}_2  - {\mathbf{x}}_3) } \cr
&\times&
\langle
\mbox{Tr} [G_{t_1}^{\dag}(\mathbf{x}_2,t_2)G_{t_1}(\mathbf{x}_2,t_2)] \cr
&\times&
\mbox{Tr} [G_{t_4}^{\dag}(\mathbf{x}_3,t_3)G_{t_4}(\mathbf{x}_3,t_3)]
\rangle .
\end{eqnarray}
Let denote that
\begin{eqnarray}
\sigma(\mathbf{x}_2,t_2) & \equiv &
\mbox{Tr} [G_{t_1}^{\dag}(\mathbf{x}_2,t_2)G_{t_1}(\mathbf{x}_2,t_2)], \cr
\sigma(\mathbf{x}_3,t_3)  &\equiv&
\mbox{Tr} [G_{t_4}^{\dag}(\mathbf{x}_3,t_3)G_{t_4}(\mathbf{x}_3,t_3)],
\end{eqnarray}
then Eq.~(\ref{app:dcr}) can be rewritten as
\begin{equation}
C_{\pi\pi}^V({\mathbf p},t_4,t_3,t_2,t_1)\hspace{-0.1cm} =\hspace{-0.1cm}
\mbox{Re} \hspace{-0.2cm}\sum_{\mathbf{x}_2,\mathbf{x}_3 }\hspace{-0.1cm}
e^{ i{\mathbf p} \cdot ({\mathbf{x}}_2  - {\mathbf{x}}_3) }
\langle\sigma(\mathbf{x}_2,t_2) \sigma(\mathbf{x}_3,t_3)\rangle , \nonumber
\end{equation}
If we define the Fourier transform:
$$
\sigma( {\bf p},t )  =
\sum_{\bf x} \sigma({\bf x},t) e^{ -i{\bf p} \cdot {\bf x} },
$$
then Eq.~(\ref{app:dcr}) can be recast as
\begin{equation}
C_{\pi\pi}^V({\mathbf p},t) =
\frac{1}{T} \mbox{Re}  \sum_{t_2=0}^{T-1}
\langle\sigma(-{\mathbf p},t_2) \sigma({\mathbf p}, t_3)\rangle,
\end{equation}
where $t=t_3-t_1$, and we sum over all the time slice to improve the statistics.
Note that $t_2=t_1+1$ and  $t_4=t_3+1$.
Of course, in the center-of-mass frame,
the vacuum diagram is still accompanied by a vacuum subtraction~\cite{Fu:2012gf}.

\section{The NLO $\pi\pi$ scattering amplitude}
\label{app:ChPT NLO}
In this Appendix, we follow the original derivations and
notations in Refs.~\cite{Gasser:1983yg,Bijnens:1995yn,Bijnens:1997vq,Colangelo:2001df,Beane:2011sc}
to derive the isospin-$0$ $\pi\pi$ partial scattering amplitude in $\chi$PT at NLO.

For the elastic $\pi\pi$ scattering, the Mandelstam variables are written
in units of  physical pion mass squared $m_\pi^2$  as
$$
s = 4+\frac{4k^2}{m_\pi^2}, \,
u = -\frac{2k^2}{m_\pi^2}(1+\cos \theta), \,
t = -\frac{2k^2}{m_\pi^2}(1-\cos \theta),
$$
where $\theta$ is the scattering angle, and $k=|{\bf k}|$
is the magnitude of  the center-of-mass three-momentum of each pion.
Note that the Legendre polynomials $P_0(\cos\theta)=1$ and
$P_2(\cos\theta) = 3/2\cos^2\theta-1$.

To calculate the amplitudes on the isospin-$0$ $\pi\pi$ scattering,
one expands the combinations with $I=0$ in the $s$-channel as
$$T^0(s,t) = 3A(s,t,u) + A(t,u,s) + A(u,s,t) .$$
Using the Formula (4.14) and the notations in Ref.~\cite{Bijnens:1997vq}, we arrive at
\begin{eqnarray}
T^0(s,t) &=& x_2   \left[ 7 + 8\frac{k^2}{m_\pi^2}\right]
           + x_2^2 \left[ 5b_1 + 12b_2+48b_3+32b_4 \right] \cr
         &+& x_2^2 \frac{k^2}{m_\pi^2} 8(b_2 + 12b_3+12b_4) \cr
         &+& x_2^2 \frac{k^4}{m_\pi^4}\left[\frac{8}{3}(22b_3 + 34b_4)
                                            \right] \cr
         &+& x_2^2 \frac{k^4}{m_\pi^4}\left[\frac{16}{3}P_2(\cos\theta)(b_3 + 7b_4)) \right] \cr
         &+& x_2^2 [3F^{(1)}(s)  + G^{(1)}(t,s)  + G^{(1)}(u,s) \cr
         &+&  F^{(1)}(t) + G^{(1)}(u,t) + 3G^{(1)}(s,t) \cr
         &+& F^{(1)}(u) + G^{(1)}(t,u) + 3G^{(1)}(s,u)],
\end{eqnarray}
where $F$ and $G$ are loop integrals denoted in Ref.~\cite{Bijnens:1995yn},
and $x_2\equiv 2m_\pi^2/f_\pi^2$.
It is trivial to show that
\begin{equation}
3F^{(1)}(s)+ G^{(1)}(t,s) + G^{(1)}(u,s)\hspace{-0.1cm} =
\hspace{-0.1cm}\bar{J}(s)\hspace{-0.1cm}\left[\frac{49}{2}
 \hspace{-0.05cm}+\hspace{-0.05cm} \frac{56k^2}{m_\pi^2}\hspace{-0.05cm}+\hspace{-0.05cm} \frac{32k^4}{m_\pi^4}\right],
\nonumber
\end{equation}
and
\begin{eqnarray}
F^{(1)}(t) + G^{(1)}(u,t) + 3G^{(1)}(s,t) &=& \frac{1}{6}\bar{J}(t)(C_{ut} - D_{ut}) , \cr
F^{(1)}(u) + G^{(1)}(t,u) + 3G^{(1)}(s,u) &=& \frac{1}{6}\bar{J}(u)(C_{ut} + D_{ut}) ,
\nonumber
\end{eqnarray}
where for easy notation, we define
\begin{eqnarray}
C_{ut} &\equiv& 5 + \frac{16 k^2}{m_\pi^2} + \frac{24 k^4}{m_\pi^4}
 + \frac{40 k^4}{m_\pi^4}\cos^2\theta , \cr
D_{ut} &\equiv& 16\cos\theta\frac{k^2}{m_\pi^2}\left(3 + \frac{4k^2}{m_\pi^2}\right),
\end{eqnarray}
and the loop function $\bar{J}(s)$ introduced Ref.~\cite{Gasser:1983yg}
is formally denoted as
\begin{equation}
\bar{J}(s) = \frac{1}{16\pi^2}\sqrt{z}\log \left( \frac{\sqrt{z}-1}{\sqrt{z}+1} \right) + \frac{1}{8\pi^2},
\quad z=1-\frac{4}{s}. \nonumber
\end{equation}
For later simple notation, we denote
\begin{equation}
K(s) \equiv \frac{1}{16\pi^2}\sqrt{z}\log \left( \frac{\sqrt{z}-1}{\sqrt{z}+1} \right) . \nonumber
\end{equation}

Using the power representations of the loop integrals (B.1) in Ref~\cite{Colangelo:2001df},
we can write
\begin{eqnarray}
\bar{J}(u) &=&  \frac{1}{96\pi^2}\left( u + \frac{u^2}{10} + \frac{u^3}{70} + {\cal{O}}(u^4) \right), \cr
\bar{J}(t) &=&  \frac{1}{96\pi^2}\left( t + \frac{t^2}{10} + \frac{t^3}{70} + {\cal{O}}(t^4) \right), \nonumber
\end{eqnarray}
where we consider that the values of $u$ and $t$ are small due to the small $k^2$ value.
Note that
\begin{eqnarray}
\bar{J}(u) + \bar{J}(t) &=& \frac{1}{96\pi^2}\left( u+t + \frac{(u+t)^2}{10} + \frac{(u+t)^3)}{70} \right) \cr
&-&\frac{1}{96\pi^2}\left( \frac{2ut}{10} + \frac{3ut(u+t)}{70} \right) + \cdots  \cr
&=& \bar{J}(u+t) - \frac{1}{96\pi^2}\hspace{-0.1cm}\left( \frac{2ut}{10} + \frac{3ut(u+t)}{70} \right)\hspace{-0.1cm} + \cdots, \cr
J(u) - J(t)  &=& \frac{u-t}{96\pi^2}\hspace{-0.1cm}\left(1 + \frac{u+t}{10} + \frac{u^2+ut+t^2}{70} \right)\hspace{-0.1cm} + \cdots.  \nonumber
\end{eqnarray}
Now it is ready to show that
\begin{eqnarray}
\bar{J}(u) + \bar{J}(t) &=& \bar{J}(-\frac{4k^2}{m_\pi^2}) \cr
            &-& \frac{1}{120\pi^2} \frac{k^4}{m_\pi^4}(1- \cos^2\theta) \left( 1
- \frac{6k^2}{7m_\pi^2} \right) + {\cal{O}}(k^6)\,, \cr
\hspace{-0.3cm}\bar{J}(u) - \bar{J}(t) &=& -\frac{1}{24\pi^2}\frac{k^2}{m_\pi^2}\cos\theta \cr
            &\times&\hspace{-0.1cm}\left(\hspace{-0.1cm} 1 - \frac{2k^2}{5m_\pi^2} + \frac{2k^4}{35m_\pi^4}(3 \hspace{-0.1cm}+\hspace{-0.1cm} \cos^2\theta) \hspace{-0.1cm}\right)\hspace{-0.1cm} + {\cal{O}}(k^6) . \nonumber
\end{eqnarray}
Hence, it is obvious to show that
\begin{eqnarray}
\hspace{-0.3cm}&&F^{(1)}(t) + G^{(1)}(u,t) + 3G^{(1)}(s,t) +\cr
\hspace{-0.3cm}&& F^{(1)}(u) + G^{(1)}(t,u) + 3G^{(1)}(s,u) \cr
\hspace{-0.3cm}&=& \frac{1}{6}\left[(\bar{J}(u)+\bar{J}(t))C_{ut}
                   + (\bar{J}(u)-\bar{J}(t))D_{ut}\right] \cr
\hspace{-0.3cm}&=& \bar{J}\left(-\frac{4k^2}{m_\pi^2}\right)
\hspace{-0.1cm}\left[\frac{5}{6} + \frac{8}{3}\frac{k^2}{m_\pi^2} + \frac{56}{9}\frac{k^4}{m_\pi^4}
        + \frac{40}{9}\frac{k^4}{m_\pi^4} P_2(\cos\theta)\right] \cr
\hspace{-0.3cm}&-&\frac{25}{216\pi^2} \frac{k^4}{m_\pi^4} -
\frac{47}{216\pi^2} \frac{k^4}{m_\pi^4} P_2(\cos\theta) + {\cal{O}}(k^6) .
\label{appb:FG}
\end{eqnarray}
We thus arrive at
\begin{eqnarray}
\label{appb:T0}
\hspace{-0.5cm}T^0(s,t) &=& x_2\left[ 7 + \frac{8k^2}{m_\pi^2} \right] \cr
\hspace{-0.5cm}         &+& x_2^2 \frac{1}{16\pi^2} [ 49+5\bar{b}_1 + 12\bar{b}_2+48\bar{b}_3+32\bar{b}_4] \cr
\hspace{-0.5cm}         &+& x_2^2\frac{k^2}{16\pi^2 m_\pi^2}\left[ \frac{352}{3} + 8(\bar{b}_2 + 12\bar{b}_3+12\bar{b}_4) \right] \cr
\hspace{-0.5cm}         &+&  x_2^2  \frac{k^4}{16\pi^2 m_\pi^4}\left[ \frac{2014}{27} + \frac{8}{3}(22\bar{b}_3 + 34\bar{b}_4)\right] \cr
\hspace{-0.5cm}         &+& x_2^2 K\hspace{-0.1cm}\left(4+\frac{4k^2}{m_\pi^2}\right)\hspace{-0.1cm}
         \left[\frac{49}{2} + 56\frac{k^2}{m_\pi^2} + 32\frac{k^4}{m_\pi^4}\right] \cr
\hspace{-0.5cm}         &+& x_2^2 K\hspace{-0.1cm}\left(-\frac{4k^2}{m_\pi^2}\right)\hspace{-0.1cm}
         \left[\frac{5}{6} + \frac{8}{3}\frac{k^2}{m_\pi^2} + \frac{56}{9}\frac{k^4}{m_\pi^4} \right] \cr
\hspace{-0.5cm}         &+&x_2^2\frac{k^4}{m_\pi^4}P_2(\cos\theta)\bigg[
            -\frac{47}{216\pi^2}  + \frac{1}{3\pi^2} (\bar{b}_3 + 7\bar{b}_4)  \cr
\hspace{-0.5cm}         && +  \frac{40}{9} \bar{J}\left(-\frac{4k^2}{m_\pi^2}\right) \bigg] .
\end{eqnarray}
where we ignore the ${\cal{O}}(k^6)$ in Eq.~(\ref{appb:FG})
since we are only interested in the terms up to and
including the ${\cal{O}}(k^4)$ term in the end.

To compare the theoretical amplitudes with data on
the isospin-$0$ $\pi\pi$ scattering ({\it e.g.}, lattice data, {\it etc}),
it is traditional to expand the combination with isospin-$0$ in the $s$-channel
$T^0(s,t)$ into the partial waves,
\begin{equation}
T^0(s,t) = 32\pi\sum_{\ell=0}^{\infty}(2\ell+1)P_\ell(\cos\theta)t_{\ell}^{0}(s) .
\end{equation}
Consequently, we get the partial wave for $s$-wave ($\ell=0$)
\begin{eqnarray}
t_0^0(s,t) &=& \frac{m_\pi^2}{16\pi f_\pi^2}\left[ 7 + 8\frac{k^2}{m_\pi^2} \right] \cr
         &+&  \frac{m_\pi^4}{128 \pi^3 f_\pi^4} \left[ 49+5\bar{b}_1 + 12\bar{b}_2+48\bar{b}_3+32\bar{b}_4 + \frac{5}{3} \right] \cr
         &+& \frac{m_\pi^2}{64 \pi^3 f_\pi^4}k^2\left[ \frac{281}{9} + 4(\bar{b}_2 + 12\bar{b}_3+12\bar{b}_4) + \frac{247}{9} \right] \cr
         &+& \frac{1}{24 \pi^3 f_\pi^4} k^4\left[ -\frac{253}{72} + 11\bar{b}_3 + 17\bar{b}_4 + \frac{35}{2} \right] \cr
&+& \frac{1}{\pi f_\pi^4} K\hspace{-0.1cm}\left(4+\frac{4k^2}{m_\pi^2}\right)\hspace{-0.1cm}
         \left[\frac{49}{16} m_\pi^4  + 7 m_\pi^2 k^2 + 4 k^4 \right] \cr
          &+& \frac{1}{\pi f_\pi^4} K\hspace{-0.1cm}\left(-\frac{4k^2}{m_\pi^2}\right)\hspace{-0.1cm}
         \left[\frac{5}{48} m_\pi^4 + \frac{1}{3} m_\pi^2 k^2 + \frac{7}{9} k^4 \right],
\label{app:t00}
\end{eqnarray}
and that for the $d$-wave ($\ell=2$)
\begin{equation}
\frac{t_2^0(s,t)}{k^4}  \hspace{-0.01cm} = \hspace{-0.01cm} \frac{1}{120\pi^3 f_\pi^4}
 \hspace{-0.02cm}\left(\hspace{-0.02cm} -\frac{47}{72}  + \bar{b}_3 + 7\bar{b}_4\hspace{-0.02cm} \right)
 + \frac{1}{36 \pi f_\pi^4} J(-\frac{4k^2}{m_\pi^2})  ,
\end{equation}
where the convention of $f_\pi$ is about $130$~MeV.
We should remark that three constants: $5/3$, $247/9$,
and $35/2$ in the right hand of Eq.~(\ref{app:t00})
will be neatly cancelled out in the Taylor expansion of
last two terms in Eq.~(\ref{app:t00}) for small $k^2$.

The near threshold (to be specific, $k^2\rightarrow 0$) behavior for
the real part of the partial wave amplitude $t(k)$
can be normally expressed as a power-series expansion in the center-of-mass energy
\begin{equation}
\mbox{Re}\; t_\ell^I(k)\ = k^{2\ell}\left\{a_\ell^I +k^2\;b_\ell^I + k^4\;c_\ell^I + {\cal O}(k^6)\right\},
\label{app:threshexp}
\end{equation}
where the threshold parameters $a_\ell^I$ and $b_\ell^I$ are referred to as the scattering lengths and  slope parameters, respectively. Of course, $c_\ell^I$ is naturally regarded as an another slope parameter.
Matching Eq.~(\ref{app:threshexp}) to  Eq.~(\ref{app:t00}) yields
\begin{eqnarray}
a_0^0 &=& \frac{7m_\pi^2}{16\pi f_\pi^2}\hspace{-0.02cm}\left[ 1 +\hspace{-0.02cm}
            \frac{m_\pi^2}{56\pi^2 f_\pi^2} \left(49\hspace{-0.01cm}+\hspace{-0.01cm}5\bar{b}_1
            \hspace{-0.01cm}+\hspace{-0.01cm} 12\bar{b}_2\hspace{-0.01cm}+\hspace{-0.01cm}48\bar{b}_3\hspace{-0.01cm}+\hspace{-0.01cm}32\bar{b}_4 \right)\right] , \cr
b_0^0 &=&  \frac{1}{2\pi f_\pi^2} \left[ 1 +
           \frac{m_\pi^2}{32 \pi^2 f_\pi^2}\left( \frac{281}{9} + 4\bar{b}_2 + 48\bar{b}_3+48\bar{b}_4  \right) \right],\cr
c_0^0 &=&  \frac{1}{24 \pi^3 f_\pi^4} \left[ -\frac{253}{72} + 11\bar{b}_3 + 17\bar{b}_4) \right], \cr
a_2^0 &=& \frac{1}{120\pi^3 f_\pi^4} \left( -\frac{47}{72}  + \bar{b}_3 + 7\bar{b}_4\right)  .
\label{app:abc}
\end{eqnarray}

It is obvious that we  nicely reproduce the relevant results in Appendix C of Ref.~\cite{Bijnens:1997vq}
for the scattering length $a_\ell^I$  and the slope parameter $b_\ell^I$ at NLO, as expected.
Note that we also give the explicit formula for the slope parameter $c_0^0$ in Eq.~(\ref{app:abc})
using the CGL language in Ref.~\cite{Bijnens:1997vq}.
In this appendix, to double-check our relevant results for the $s$-wave,
we also list the results for $d$-wave,
since the scattering amplitude in Eq.~(\ref{appb:T0}) includes both information of them.


\end{document}